\documentclass[aps,prd,final,nofootinbib,superscriptaddress]{revtex4}
\pdfoutput=1
\usepackage{amssymb,amsmath,amstext,amsfonts}
\usepackage{bbold,ulem,bm}
\usepackage{graphics}
\usepackage{mathtools}
\usepackage{hyperref}
\usepackage{color}

\usepackage{graphicx}
\usepackage{mwe}
\usepackage{mathrsfs}

\newcommand{\beq}{\begin{equation}}
\newcommand{\eeq}{\end{equation}}
\newcommand{\ba}{\begin{eqnarray}}
\newcommand{\ea}{\end{eqnarray}}
\newcommand{\bal}{\begin{aligned}}
\newcommand{\eal}{\end{aligned}}
\newcommand{\Lag}{{\mathcal{L}}}

\begin{document}

\title{Degeneracy, matter coupling, and disformal transformations\\
 in scalar-tensor theories} 

\author{C\'edric Deffayet,}
\affiliation{Sorbonne Universit\'e, UPMC Paris 6 and CNRS, UMR 7095, Institut d'Astrophysique de Paris, GReCO, 98bis boulevard Arago, 75014 Paris, France}
\affiliation{IHES, Le Bois-Marie, 35 route de Chartres, 91440 Bures-sur-Yvette, France}
\author{Sebastian Garcia-Saenz}
\affiliation{Theoretical Physics, Blackett Laboratory, Imperial College, London, SW7 2AZ, U.K.}

\begin{abstract}
Degenerate scalar-tensor theories of gravity extend general relativity by a single degree of freedom, despite their equations of motion being higher than second order. In some cases, this is a mere consequence of a disformal field redefinition carried out in a non-degenerate theory. More generally, this is made possible by the existence of an additional constraint that removes the would-be ghost.  It has been noted that this constraint can be thwarted when the coupling to matter involves time derivatives of the metric, which results in a modification of the canonical momenta of the gravitational sector. In this note we expand on this issue by analyzing the precise ways in which the extra degree of freedom may reappear upon minimal coupling to matter. Specifically, we study examples of matter sectors that lead either to a direct loss of the special constraint or to a failure to generate a pair of secondary constraints. We also discuss the recurrence of the extra degree of freedom using the language of disformal transformations in particular for what concerns ``veiled" gravity. On the positive side, we show that the minimal coupling of spinor fields is healthy and does not spoil the additional constraint. We argue that this virtue of spinor fields to preserve the number of degrees of freedom in the presence of higher derivatives is actually very general and can be seen from the level decomposition of Grassmann-valued classical variables.
\end{abstract}


\maketitle


\section{Introduction} \label{sec:intro}

Scalar-tensor theories of gravity are appealing for a number of reasons \cite{Fujii:2003pa,Quiros:2019ktw}. The resounding experimental success of general relativity (GR) suggests that if gravity is to be modified in the infrared, we had better do so in a conservative way, and the most minimal tweak to be done is to add a single scalar degree of freedom besides the graviton described by GR. At the same time, we may hope that such a minimal field content would allow for a strong theoretical control, permitting for instance to accomplish a thorough classification of scalar-tensor models of gravity. 

This effort of charting the space of all theories describing the dynamics of a single spin-0 and a single massless spin-2 particles has indeed been an active research program over the past decade, beginning with the rediscovery of Horndeski theory \cite{Horndeski:1974wa,Deffayet:2009mn} (following that of the galileon \cite{Nicolis:2008in} and its covariant version \cite{Deffayet:2009wt}). While Horndeski theory comprises the most general action leading to manifestly second-order equations of motion for both the scalar field and the metric tensor, hence ensuring the correct number of degrees of freedom (DoF) classically, it is also clear that such a theory (or in fact mere general relativity) can be ``disguised" into a higher derivative theory via an invertible field redefinition such as the well-known disformal transformations first considered by Bekenstein \cite{Bekenstein:1992pj}. More generally it was realized that an action can produce higher-order eqs.\ of motion while still giving rise to the desired number of three DoF, thanks to the existence of degeneracies among the equations so that the required pieces of initial data are reduced. In the context of the Hamilton--Dirac analysis, these degeneracies manifest themselves as additional pairs of second class constraints, thus making very transparent that some of the DoF that one would naively infer from the action are actually non-dynamical \cite{Zumalacarregui:2013pma,Lin:2014jga,Gleyzes:2014dya,Deffayet:2015qwa,Langlois:2015skt,Motohashi:2016ftl,Klein:2016aiq,deRham:2016wji}. This leads to the generalization of Horndeski theories to ``beyond Horndeski'' \cite{Zumalacarregui:2013pma,Gleyzes:2014dya,Gleyzes:2014qga,Langlois:2015cwa,Crisostomi:2016tcp} and eventually to the larger class of degenerate higher-order scalar-tensor theories (DHOST) \cite{Crisostomi:2016czh,Achour:2016rkg,BenAchour:2016fzp} (see \cite{Deffayet:2013lga,Langlois:2018dxi,Kobayashi:2019hrl} for reviews).

Although this story is by now well understood, a rather unexplored question concerns the coupling to matter in DHOST. Applications of DHOST coupled to certain matter fields have of course been considered, particularly in the contexts of cosmology and astrophysics (see e.g.\ \cite{Berezhiani:2013dw,Babichev:2016rlq,Langlois:2017dyl,Babichev:2018rfj,Crisostomi:2018bsp,Hirano:2019nkz,Hirano:2019scf,Crisostomi:2019yfo,BenAchour:2019fdf}), yet a general understanding of the consistency of matter coupling is currently lacking. This is in fact rather surprising: indeed, besides its obvious relevance for phenomenology, this question is crucial for the claim that the considered theory propagates fewer degrees of freedom than naively expected. Consider for example the class of DHOST theories whose extended gravity (i.e.\ metric and scalar) sector can be obtained explicitly out of non-degenerate scalar-tensor theories via a field redefinition. Obviously, these theories, in the absence of matter, are just the same as their non-degenerate counterparts. The only difference between them can only be coming from the coupling to matter, which in turn can spoil the degeneracy (some examples are given below) and the equivalence between the considered theories. This is all the more true if the field redefinition is disformal, as it involves in this case a derivative of the scalar which can potentially lead to a kinetic mixing with matter degrees of freedom. 

The consistency of matter coupling is obviously a very broad question that is hard to address in full generality, but an interesting first step is to define such a consistency problem as follows: {\it we shall say that a matter field can be coupled consistently to DHOST if the minimal coupling prescription preserves the number of degrees of freedom of the extended gravity sector in the absence of matter}. In other words, the complete counting should yield three DoF for the scalar-tensor sector plus whatever number of DoF the matter sector had in the absence of gravity. It is important to emphasize that this is already a non-trivial question in pure GR, even for matter theories without gauge invariance, the reason being that minimal gravitational coupling can spoil some of the constraints that a matter action would otherwise have in flat spacetime \cite{Isenberg:1977fs}. 

Our goal in this note is to point out that this issue is even more delicate in DHOST. The reason is quite simple: gravitational coupling in DHOST is not only a threat to the constraints of the matter sector, but it can also doom the special constraint that ensures the degeneracy of the scalar-tensor eqs.\ of motion and is responsible for removing the Ostrogradski ghost. We remark that this problem was already observed in \cite{deRham:2016wji}, where it was explained that the presence of time derivatives of the lapse function, which in pure DHOST can always be removed via a field redefinition, can become truly pathological when coupled to additional fields. We seek in this note to further clarify this aspect through a full Hamilton--Dirac analysis of three instructive examples of matter fields coupled to a generic quadratic DHOST theory. Although of course this does not encompass the most general set of scalar-tensor models, it should be clear that the general lessons we will draw should apply very generically. The case of quadratic DHOST theory is also worth focusing on given its interest in the astrophysical and cosmological literature (see e.g.\ \cite{Babichev:2017guv,Bartolo:2017ibw,DeFelice:2018mkq,Santoni:2018rrx,Frusciante:2018tvu,Motohashi:2019sen,Charmousis:2019vnf,deRham:2019slh,Minamitsuji:2019shy}). Our analysis also opens the way to a classification based on matter coupling consistencies (in the above terminology) of the different scalar-tensor theories among themselves as well as compared with pure general relativity.

These issues indeed also arise in the case of pure gravity disguised in the form of the so-called ``veiled" gravity via a disformal transformation. Indeed, it will be shown that the mere minimal coupling of veiled gravity to a single scalar is enough to make dynamical the scalar appearing in the disformal transformation. This serves as an illustrating simple starting point for the rest of the discussion which uses a Hamiltonian approach. The first of our examples is then a toy non-canonical vector model that leads to a loss of the DHOST primary constraint and hence to the reappearance of the Ostrogradski ghost. The second case is the cubic galileon considered in \cite{deRham:2016wji}, which spoils the DoF count in the Hamilton--Dirac analysis in a more subtle way, namely by preventing the primary constraints of the DHOST and galileon sectors to generate their corresponding secondary constraints. Lastly we consider the physically relevant example of a Majorana spinor field coupled to DHOST, which we will show to be consistent in the sense defined above despite what one might naively think at first given the higher order nature of the field equations. We will argue however that this positive result is not an accident of DHOST but rather a generic virtue of classical spinor fields: higher-order derivatives of a field coupled to a spinor are in many cases harmless (at least ``classically" and from the point of view of the Hamilton--Dirac counting of DoF) as a result of the so-called level decomposition of Grassmann-valued variables. Such harmless couplings include e.g.\ curvature dependent mass-terms of the form $R^n\bar{\lambda}\lambda$ which can have an interesting phenomenology.

To outline the rest of the paper, in sec.\ \ref{sec:dhost} we briefly review the formulation of quadratic DHOST models that we will focus on, including the Hamiltonian analysis of \cite{Langlois:2015skt}. We also discuss there the simple case of veiled gravity. In sec.\ \ref{sec:matter} we consider the coupling to matter in DHOST, and show through two examples of matter fields the possible ways in which minimal gravitational coupling can render the theory inconsistent according the aforementioned criterion. We give a separate treatment of the coupling of spinors to DHOST in sec.\ \ref{sec:spinors}, focusing on a minimally coupled Majorana spinor. We end our note in sec.\ \ref{sec:discussion} with some general conclusions and comments.

\vspace{1em}

\noindent
{\it Conventions:} We work in four spacetime dimensions and use the mostly plus signature for the metric. Greek indices stand for spacetime coordinates ($\mu,\nu,\ldots=0,1,2,3$), latin indices for spatial coordinates ($i,j,\ldots=1,2,3$). Symmetrizations and anti-symmetrizations of indices are defined with unit weight. When we deal with spinors in sec.\ \ref{sec:spinors} we will need to further distinguish the tangent space coordinates, which we denote with latin indices starting with $a,b,\ldots=0,1,2,3$, as well as 4-component spinor indices that we denote with greek indices starting with $\alpha,\beta,\ldots=1,2,3,4$. See footnote \ref{fn:spinor conventions} for more explanations on our conventions for spinors.


\section{Degenerate higher-order scalar-tensor theories} \label{sec:dhost}

The general quadratic DHOST gravitational theory for a metric $g_{\mu\nu}$ and scalar field $\phi$ is defined by the action \cite{Langlois:2015cwa}
\beq \label{eq:full dhost action}
S_g[g,\phi]=\int d^4x\sqrt{-g}\Big[F(\phi,X)R+P(\phi,X)+Q(\phi,X)\Box\phi+C^{\mu\nu\rho\sigma}[\phi]\nabla_{\mu}\nabla_{\nu}\phi\nabla_{\rho}\nabla_{\sigma}\phi\Big]\,,
\eeq
where $R$ is the 4-dimensional curvature scalar, $X:=\nabla^{\mu}\phi\nabla_{\mu}\phi$ and
\beq\bal
C^{\mu\nu\rho\sigma}&:=A_1g^{\mu(\rho}g^{\sigma)\nu}+A_2g^{\mu\nu}g^{\rho\sigma}+\frac{A_3}{2}\left(\phi^{\mu}\phi^{\nu}g^{\rho\sigma}+\phi^{\rho}\phi^{\sigma}g^{\mu\nu}\right)\\
&\quad+\frac{A_4}{2}\left(\phi^{\mu}\phi^{(\rho}g^{\sigma)\nu}+\phi^{\nu}\phi^{(\rho}g^{\sigma)\mu}\right)+A_5\phi^{\mu}\phi^{\nu}\phi^{\rho}\phi^{\sigma}\,,
\eal\eeq
with $\phi_{\mu}:=\nabla_{\mu}\phi$ and the $A$'s are functions of $\phi$ and $X$. Propagation of three DoF, and hence absence of an Ostrogradski ghost, imposes certain constraints among the functions $A$'s. In addition, there exist some subclasses of degenerate theories which also constrain the function $F$, while $P$ and $Q$ are always arbitrary as far the counting of DoF is concerned. For our purposes we will not need the precise form of these relations among the various functions (the reader may find them in \cite{Langlois:2018dxi}), but simply assume the existence of a degeneracy. We will also ignore special cases with even more degeneracies that lead to less than three DoF, except when we will discuss the case of ``veiled" gravity in subsection \ref{2.2}.

For the Hamilton--Dirac analysis it is convenient to first eliminate the second derivatives of the scalar field by introducing an auxiliary vector $A_{\mu}$, with the relation $A_{\mu}=\nabla_{\mu}\phi$ being enforced by a Lagrange multiplier \cite{Deffayet:2015qwa,Langlois:2015cwa,Langlois:2015skt}. Thus we consider the modified action
\beq\bal \label{eq:first order dhost action}
S_g[g,\phi,A,\lambda]&=\int d^4x\Big\{\sqrt{-g}\Big[F(\phi,X)R+P(\phi,X)+Q(\phi,X)\nabla_{\mu}A^{\mu}\\
&\quad+C^{\mu\nu\rho\sigma}[\phi,A]\nabla_{\mu}A_{\nu}\nabla_{\rho}A_{\sigma}\Big]+\lambda^{\mu}(A_{\mu}-\nabla_{\mu}\phi)\Big\}\,,
\eal\eeq
and it is understood that now $X=A^{\mu}A_{\mu}$ and likewise all instances of $\phi_{\mu}$ in $C^{\mu\nu\rho\sigma}$ have been replaced.

The next step is to introduce ADM variables for the metric \cite{Arnowitt:1962hi},
\beq
g_{\mu\nu}dx^{\mu}dx^{\nu}=-N^2dt^2+\gamma_{ij}(dx^i+N^idt)(dx^j+N^jdt)\,,
\eeq
and perform a $3+1$ decomposition of every operator in the action. The first three terms in \eqref{eq:first order dhost action} (which we refer to as the KGB action following \cite{Deffayet:2010qz}) yield\footnote{We raise and lower latin indices with the 3-metric $\gamma_{ij}$, for instance $A^i=\gamma^{ij}A_j$. The covariant derivative compatible with $\gamma_{ij}$ is denoted by $D_i$.}
\beq
S_{\rm KGB}=\int dtd^3xN\sqrt{\gamma}\Big\{F\Big[K^{ij}K_{ij}-K^2+R^{(3)}-2\nabla_{\mu}(a^{\mu}-Kn^{\mu})\Big]+P+Q\nabla_{\mu}A^{\mu}\Big\}\,,
\eeq
where $K_{ij}$ is the extrinsic curvature, $K:=\gamma^{ij}K_{ij}$ and $R^{(3)}$ is the curvature scalar associated to the 3-metric. As explained in \cite{Langlois:2015skt}, the presence of the function $F$ means that there is an extra contribution relative to the standard Einstein--Hilbert result which involves the vectors
\beq
n^{\mu}:=\frac{1}{N}\left(1,-N^i\right)\,,\qquad a^{\mu}:=n^{\nu}\nabla_{\nu}n^{\mu}\,.
\eeq
Here $n^{\mu}$ corresponds to the vector normal to the constant-time hypersurfaces (it is normalized, $n^{\mu}n_{\mu}=-1$) and $a^{\mu}$ is the ``acceleration'' of the integral curves of $n^{\mu}$ (see e.g.\ \cite{Carroll:2004st}). In ADM components one has $a^{\mu}=(0,D^iN/N)$.

Completing the $3+1$ decomposition one finds the result of \cite{Langlois:2015skt}, here slightly generalized to include the KGB terms:\footnote{It helps in the calculation to know that $\nabla_{\mu}n^{\mu}=K$ and $\nabla_{\mu}a^{\mu}=\frac{1}{N}D^iD_iN$.}
\beq\bal \label{eq:3+1 KGB}
S_{\rm KGB}&=\int dtd^3xN\sqrt{\gamma}\bigg[\frac{2}{N}\,{\cal B}^{ij}_{\rm KGB}K_{ij}(\dot{A}_{*}-\Xi_A)+{\cal K}^{ij,kl}_{\rm KGB}K_{ij}K_{kl}+2{\cal C}^{ij}_{\rm KGB}K_{ij}\\
&\quad+\frac{2}{N}\,{\cal C}^0_{\rm KGB}(\dot{A}_{*}-\Xi_A)-{\cal U}_{\rm KGB}\bigg]\,,
\eal\eeq
where
\beq \label{eq:Astar def}
A_{*}:=n^{\mu}A_{\mu}=\frac{1}{N}(A_0-N^iA_i)\,,\qquad \Xi_A:=A^iD_iN+N^iD_iA_{*}\,,
\eeq
and
\beq\bal \label{eq:KGB tensors}
{\cal B}^{ij}_{\rm KGB}&=2F_XA_{*}\gamma^{ij}\,,\\
{\cal K}^{ij,kl}_{\rm KGB}&=F\left(\gamma^{i(k}\gamma^{l)j}-\gamma^{ij}\gamma^{kl}\right)+2F_X\left(\gamma^{ij}A^kA^l+\gamma^{kl}A^iA^j\right)\,,\\
{\cal C}^{ij}_{\rm KGB}&=-(F_{\phi}A_{*}+2F_XA^iD_iA_{*})\gamma^{ij}-\frac{1}{2}\,QA_{*}\gamma^{ij}\,,\\
{\cal C}^0_{\rm KGB}&=-\frac{1}{2}\,Q\,,\\
{\cal U}_{\rm KGB}&=-R^{(3)}+2D^iD_iF-P-QD_iA^i\,,
\eal\eeq
with $F_{\phi}:=\partial F/\partial \phi$ and $F_X:=\partial F/\partial X$. Note that $A_{*}$ is to be regarded as a dynamical variable instead of $A_0$, so that the set of tensors in \eqref{eq:KGB tensors} are independent of the lapse and shift.\footnote{In particular $X=-A_{*}^2+A^iA_i$ is independent of $N$ and $N^i$.} They also do not involve any time derivatives, which only appear in the form of $K_{ij}$ and $\dot{A}_{*}$, while time derivatives of $A_i$ can always be removed by employing the Lagrange constraint.

To the decomposed KGB action in \eqref{eq:3+1 KGB} we must add the quadratic DHOST contribution
\beq
S_{\rm quad}=\int d^4x\sqrt{-g}C^{\mu\nu\rho\sigma}[\phi,A]\nabla_{\mu}A_{\nu}\nabla_{\rho}A_{\sigma}\,.
\eeq
The resulting 3+1 decomposition is of the same structure as \eqref{eq:3+1 KGB} but with an additional term that is quadratic in $\dot{A}_{*}$,
\beq\bal \label{eq:3+1 quad}
S_{\rm quad}&=\int dtd^3xN\sqrt{\gamma}\bigg[\frac{1}{N^2}\,{\cal A}(\dot{A}_{*}-\Xi_A)^2+\frac{2}{N}\,{\cal B}^{ij}_{\rm quad}K_{ij}(\dot{A}_{*}-\Xi_A)+{\cal K}^{ij,kl}_{\rm quad}K_{ij}K_{kl}\\
&\quad+2{\cal C}^{ij}_{\rm quad}K_{ij}+\frac{2}{N}\,{\cal C}^0_{\rm quad}(\dot{A}_{*}-\Xi_A)-{\cal U}_{\rm quad}\bigg]\,,
\eal\eeq
where
\beq
{\cal A}=A_1+A_2-(A_3+A_4)A_{*}^2+A_5A_{*}^4\,.
\eeq
We will actually not need this explicit expression nor those of the other tensors (which can be found in \cite{Langlois:2015skt}); it should only be remembered that they involve $\phi$, $A_{*}$, $A_i$ and $\gamma_{ij}$, but not their time derivatives or the lapse and shift.

Collecting everything we arrive at the full $3+1$-decomposed gravitational action,
\beq\bal \label{eq:3+1 full}
S_g&=\int dtd^3x\Big\{N\sqrt{\gamma}\Big[{\cal A}V_{*}^2+2{\cal B}^{ij}V_{*}K_{ij}+{\cal K}^{ij,kl}K_{ij}K_{kl}+2{\cal C}^{ij}K_{ij}+2{\cal C}^0V_{*}-{\cal U}\Big]\\
&\quad+\lambda^0\left(NA_{*}+N^iA_i-\dot{\phi}\right)+\lambda^i\left(A_i-D_i\phi\right)\Big\}\,,
\eal\eeq
where ${\cal B}^{ij}:={\cal B}^{ij}_{\rm KGB}+{\cal B}^{ij}_{\rm quad}$ and similarly for the other tensors. We also introduced the shorthand notation $V_{*}:=(\dot{A}_{*}-\Xi_A)/N$. Again, explicit expressions will not be needed, but only the fact that in DHOST it holds that \cite{Langlois:2015cwa}
\beq \label{eq:degeneracy cond}
{\cal A}-{\cal K}^{-1}_{ij,kl}{\cal B}^{ij}{\cal B}^{kl}=0\,.
\eeq
This is the degeneracy condition that ensures the absence of the Ostrogradski ghost, as we review next in the Hamiltonian language.

\subsection{Hamiltonian and degrees of freedom}

From the action \eqref{eq:3+1 full} we derive the canonical momenta,
\beq\begin{gathered}
\pi_0:=\frac{\partial\Lag}{\partial\dot{N}}=0\,,\qquad \pi_i:=\frac{\partial\Lag}{\partial\dot{N}^i}=0\,,\\
\pi^{ij}:=\frac{\partial\Lag}{\partial\dot{\gamma}_{ij}}=\frac{1}{2N}\,\frac{\partial\Lag}{\partial K_{ij}}=\sqrt{\gamma}\left[{\cal K}^{ij,kl}K_{kl}+{\cal B}^{ij}V_{*}+{\cal C}^{ij}\right]\,,
\end{gathered}\eeq
\beq\begin{gathered}
p_{\phi}:=\frac{\partial\Lag}{\partial\dot{\phi}}=-\lambda^0\,,\qquad p^i:=\frac{\partial\Lag}{\partial\dot{A}_i}=0\,,\\
p_{*}:=\frac{\partial\Lag}{\partial\dot{A}_{*}}=\frac{1}{N}\,\frac{\partial\Lag}{\partial V_{*}}=2\sqrt{\gamma}\left[{\cal A}V_{*}+{\cal B}^{ij}K_{ij}+{\cal C}^0\right]\,.
\end{gathered}\eeq
From these expressions one obtains the following set of primary constraints\footnote{We follow the convention of \cite{Langlois:2015skt} of not regarding the Lagrange multiplier $\lambda^{\mu}$ as an independent variable in the Hamilton--Dirac analysis. In this approach one uses the relation $\lambda^0=-p_{\phi}$ directly in the action, while the Lagrange constraint enforced by $\lambda^i$ is incorporated as a primary constraint (and it is therefore not included in the base Hamiltonian below). Note that in the alternative convention, where one does see $\lambda^{\mu}$ as a phase space variable, the Lagrange constraint $\chi_i$ would appear as a secondary constraint.}
\beq
\pi_0\approx0\,,\qquad \pi_i\approx0\,,\qquad p^i\approx0\,,\qquad \chi_i:=A_i-D_i\phi\approx0\,,
\eeq
\beq \label{eq:Psi function}
\Psi:=p_{*}-2{\cal K}^{-1}_{ij,kl}\pi^{ij}{\cal B}^{kl}+2\sqrt{\gamma}\left({\cal K}^{-1}_{ij,kl}{\cal C}^{ij}{\cal B}^{kl}-{\cal C}^0\right)\approx 0\,,
\eeq
where the last constraint is a direct consequence of the degeneracy condition \eqref{eq:degeneracy cond}.

Solving for the velocities the ``base'' Hamiltonian can be shown to reduce to
\beq\bal
H_{\rm base}&=\int d^3x\left[\pi^{ij}\dot{\gamma}_{ij}+p_{*}\dot{A}_{*}+p_{\phi}\dot{\phi}-\Lag\right]= \int d^3x\left[N{\cal H}_0+N^i{\cal H}_i\right]\,,
\eal\eeq
with
\beq\bal
{\cal H}_0&=\sqrt{\gamma}\left[{\cal K}^{-1}_{ij,kl}\left(\frac{\pi^{ij}}{\sqrt{\gamma}}-{\cal C}^{ij}\right)\left(\frac{\pi^{kl}}{\sqrt{\gamma}}-{\cal C}^{kl}\right)+{\cal U}\right]+p_{\phi}A_{*}-D_i(p_{*}A^i)\,,\\
{\cal H}_i&=-2D^j\pi_{ij}+p_{\phi}A_i+p_{*}D_iA_{*}\,,
\eal\eeq
which is the form expected for a diffeomorphism invariant theory. Adding the constraints we obtain the ``augmented'' Hamiltonian\footnote{We call the Hamiltonian that includes the primary constraints ``augmented'' to distinguish it from the ``total'' Hamiltonian that incorporates all the constraints, even though we will not need to derive the latter explicitly.}
\beq
H_{\rm aug}=\int d^3x\left[N{\cal H}_0+N^i{\cal H}_i+\mu^0\pi_0+\mu^i\pi_i+\lambda^i\chi_i+\alpha_ip^i+\xi\Psi\right]\,,
\eeq
where $\mu^0$, $\mu^i$, $\lambda^i$, $\alpha_i$ and $\xi$ are Lagrange multipliers. The augmented Hamiltonian is to be used to enforce the preservation in time of the primary constraints. Each of these conditions will either (a) be automatically satisfied, or (b) determine a Lagrange multiplier, or (c) yield a secondary constraint.

The constraints $\pi_0\approx0$ and $\pi_i\approx0$ generate the secondary constraints ${\cal H}_0\approx0$ and ${\cal H}_i\approx0$, respectively. This set of 8 constraints must be first class because of the general covariance of the theory (although, as remarked in \cite{Langlois:2015skt}, showing this explicitly can be extremely cumbersome). On the other hand, time preservation of $\chi_i\approx0$ and $p^i\approx0$ determines the associated Lagrange multipliers,\footnote{In deriving these relations we freely use the weak equality $A_i\approx D_i\phi$ to simplify the constraints and the solutions for the Lagrange multipliers.}
\beq\bal \label{eq:lagrange multiplier sol}
0&=\{\chi_i,H_{\rm aug}\}=\alpha_i+D_i(NA_{*}+N^jA_j)\qquad \Rightarrow\qquad \alpha_i=-D_i(NA_{*}+N^jA_j)\,,\\
0&=\{p^i,H_{\rm aug}\}=-\lambda^i\qquad \Rightarrow\qquad \lambda^i=0\,.
\eal\eeq
Lastly, the consistency of the DHOST constraint $\Psi\approx0$ gives $\Omega:=\{\Psi,H_{\rm aug}\}=\{\Psi,H_{\rm base}\}\approx0$, where the first equality follows because $\Psi$ commutes with the other primary constraints. Thus the last relation does not involve any Lagrange multiplier and defines a secondary constraint. In the absence of further degeneracies, as we assume, the consistency of this last constraint, $\Omega\approx0$, then fixes the multiplier $\xi$ and the Hamilton--Dirac analysis ends. The total number of second class constraints is therefore $8$.

For the final counting of DoF, the number of phase space variables for the fields $(g_{\mu\nu},A_{\mu},\phi)$ is $30=15\times2$, from which we subtract $2\times8$ for the first class constraints and $8$ for the second class constraints, giving $6$ dynamical phase space variables or $3$ DoF, which is the correct result for a theory describing a graviton and a scalar field.


\section{Coupling to matter in DHOST} \label{sec:matter}

We now address the question of whether the coupling to matter in a generic DHOST theory may be inconsistent in the restricted sense we adopted in the introduction. This means that, if a given matter field has $N$ DoF in the absence of gravity, then the coupling to DHOST will be said to be inconsistent if the full action with the matter field being minimally coupled to the metric propagates strictly more than $N+3$ DoF.

It is well known that this ``continuity'' condition can be violated already in pure GR. The case of massless higher-spin theories is a prominent instance in which minimal gravitational coupling leads to a violation of gauge invariance (see \cite{Aragone:1971kh,Aragone:1979bm,Aragone:1981yn} for early works).\footnote{Of course also well known is the fact that minimal coupling of ``matter'' particles can actually be problematic in the context of electromagnetism; see e.g.\ \cite{Federbush1961,Deser:1963zzc,Berends:1979wu}.} But there are simpler examples of matter sectors with no gauge symmetries that are also inconsistent, for the reason that they possess second class constraints that are lost upon coupling to GR \cite{Isenberg:1977fs}. It is easy to see that the ``dangerous'' matter theories are the ones that, in their covariantized versions, include the Christoffel connection and hence time derivatives of the metric. Indeed, if terms involving $\dot{\gamma}_{ij}$ are absent in the matter action then the kinetic matrix is block-diagonal and the number of primary constraints, corresponding to the kernel dimension of the kinetic matrix, is preserved by the gravitational coupling. Moreover, secondary constraints are also safe in this situation because diffeomorphism invariance ensures that the GR constraints remain first class and that the matter primary constraints come in pairs with their secondary constraints (regardless of whether they are first or second class).\footnote{There remains however the logical possibility that a matter field could also have tertiary and quaternary (or even higher order) constraints that become lost due to the coupling to gravity, but we are not aware of any example of this type.} It is therefore no surprise that the most familiar matter models --- standard scalar fields, Maxwell, Proca and Yang--Mills fields --- pose no problem to the consistency of minimal coupling (the case of spinors is less trivial; we will come back to it in sec.\ \ref{sec:spinors}).

It is then natural to ask if this issue is somehow worse in DHOST. Concretely, do there exist matter fields that are consistent when coupled to GR, but inconsistent when coupled to DHOST? The answer is yes, as we will show next through two examples of matter sectors. It should be clear that our argument does not aim to rule out the class of DHOST models (our examples are admittedly somewhat contrived) but rather to extract general lessons on the problem of matter coupling in extended scalar-tensor theories of gravity, as well as to stress the differences between DHOST and standard scalar-tensor theories or GR.  We emphasize also that similar issues can arise in DHOST which propagate strictly less than 3 DoF, the canonical example of which being the so-called ``veiled" gravity. This aspect is discussed in the next subsection as an illustrative starting point and in fact, as we will show, the situation there is even worse than in generic DHOST: even the simplest minimal coupling of veiled gravity to a mere scalar spoils the free theory DoF counting.

\subsection{Matter coupling and disformal transformations}

Consider a given DHOST theory whose extended gravity sector has an action $S_g[g_{\mu \nu},\phi]$, with a minimal coupling of the matter fields, collectively denoted by $\Phi_m$, to the metric $g_{\mu \nu}$, so that the (minimally coupled) matter action can be written as $S_m[g_{\mu \nu},\Phi_m]$. We then ask if this coupling is consistent (in the terminology of our introduction), {\it i.e.}~if it preserves the number of DoF of the free theory $S_g[g_{\mu \nu},\phi]$. This question will be addressed later in the Hamiltonian framework, however an interesting light can be shed on this issue using disformal transformations, as we now show with some generality before applying this to General Relativity and veiled gravity.

\subsubsection{Disformal transformations}

It is known that a subset of DHOST theories can be mapped to Horndeski theories via a disformal transformation \cite{Zumalacarregui:2013pma, Bettoni:2013diz}. Such a transformation is defined by the following relation \cite{Bekenstein:1992pj} between two metrics $g_{\mu \nu}$, $\tilde{g}_{\mu \nu}$ and a scalar $\phi$, 
\ba \label{disdef}
g_{\mu \nu} &=& g_{\mu \nu}\left(\tilde{g}_{\mu \nu},\phi\right) \\
 &=& a(\phi,\tilde{X}) \tilde{g}_{\mu \nu} + b(\phi,\tilde{X}) \phi_\mu \phi_\nu
\ea
where $\tilde{X}$ is defined as above from the derivative of the scalar $\phi_\mu\equiv \partial_\mu \phi$ and the inverse metric $\tilde{g}^{\mu \nu}$ as $\tilde{X}=\tilde{g}^{\mu \nu}\phi_\mu \phi_\nu$. Such a transformation is generically (i.e.\ for generic functions $a$ and $b$) invertible, with an inverse of the same form  
\ba \label{disdefinv}
\tilde{g}_{\mu \nu} &=& \tilde{g}_{\mu \nu}\left(g_{\mu \nu},\phi\right) \\
 &=& \alpha(\phi,X) g_{\mu \nu} + \beta(\phi,X) \phi_\mu \phi_\nu,
\ea
where the relation between $\left[\alpha(\phi,X), \beta(\phi,X)\right]$ and $\left[a(\phi,\tilde{X}), b(\phi,\tilde{X})\right]$ can easily be found (possibly only implicitly) and does not matter here. It also implies a relation between the Christoffel symbols $\tilde{\Gamma}^\lambda_{\mu \nu}$ and $\Gamma^\lambda_{\mu \nu}$ of the two metrics of the form
\ba \label{disformalFORM}
\Gamma^\lambda_{\mu \nu} = \tilde{\Gamma}^\lambda_{\mu \nu}  + C^\lambda_{\mu \nu}
\ea
where the exact expression of $C^\lambda_{\mu \nu}$ is not important here (it can be found at numerous places, including the seminal \cite{Zumalacarregui:2013pma}), except for the fact that it depends on up to second derivatives of the scalar $\phi$. When the above (\ref{disdef}) is invertible, it is clear that the extended gravity actions $S_g[g_{\mu \nu},\phi]$ and $\tilde{S}_g[\tilde{g}_{\mu \nu},\phi]\equiv S_g[g_{\mu \nu}=g_{\mu \nu}\left(\tilde{g}_{\mu \nu},\phi \right),\phi]$ describe the same physics and in particular have the same number of DoF, even if it can be that the first one is degenerate while the second is not. Let us assume we are now in this situation. The equivalence then also holds if one adds to the first theory a minimal matter coupling $S_m[g_{\mu \nu},\Phi_m]$ and to the second, its disformally transformed one: $\tilde{S}_m[\tilde{g}_{\mu \nu},\phi ,\Phi_m] \equiv S_m[g_{\mu \nu}=g_{\mu\nu}\left(\tilde{g}_{\mu \nu},\phi \right),\Phi_m]$. The latter coupling makes however the non degenerate scalar-tensor theory $\tilde{S}_g[\tilde{g}_{\mu \nu},\phi]$ non minimally coupled to matter. This can lead in fact to an increase in the number of propagating DoF compared to the situation where the same non degenerate scalar-tensor theory $\tilde{S}_g[\tilde{g}_{\mu \nu},\phi]$ would have been minimally coupled to the very same matter fields $\Phi_m$. When this happens, it also means that the original minimal matter coupling of the original DHOST is not consistent (in the terminology of the introduction to this work). Using disformal transformations, it is easy to understand why such an increase can happen: indeed the disformal transformation (\ref{disdef}) contains a first derivative of the scalar. As a result any occurrence of the metric $g_{\mu \nu}$ in the minimal coupling $S_m[g_{\mu \nu},\Phi_m]$ will contain a first derivative of the scalar $\phi$ when expressed in the action $\tilde{S}_m[\tilde{g}_{\mu \nu},\phi ,\Phi_m]$ and any $g_{\mu \nu}-$covariant derivative of a given matter field $\Phi_m$ which appear in the minimal coupling $S_m[g_{\mu \nu},\Phi_m]$ will contain a second derivative of $\phi$ when expressed in the action $\tilde{S}_m[\tilde{g}_{\mu \nu},\phi ,\Phi_m]$ (as a consequence of the above discussion for Christoffel symbols). In general, the occurrence of first derivative of $\phi$ in the matter coupling can result in a mixing with matter and is not worrisome if this scalar already propagates in the action $\tilde{S}_g[\tilde{g}_{\mu \nu},\phi]$; it is however worrisome when this scalar does not propagate (this is precisely what happens in the example discussed in the next subsection). The occurrence of second derivatives of $\phi$ is however more worrisome in general and can lead to an inconsistent coupling. Hence we expect to find inconsistent matter coupling in theories with minimal coupling involving covariant derivatives of the matter field. Fortunately, scalars or gauge-invariant p-forms (as the latter have actions with exterior derivatives which do not involve covariant derivatives) are not of the latter type. This is not true for non gauge-invariant forms. Similarly, and more importantly, this also does not hold for fermions, and the consistency of their matter coupling appears hence worth of investigation.

\subsubsection{The example of veiled gravity}

Veiled gravity, as the name indicates, is just General Relativity disguised via a disformal transformation \cite{Deruelle:2014zza}. I.e.\ we can consider the standard Einstein Hilbert action 
\ba
\int d^4 x \sqrt{-\tilde{g}} \tilde{R}
\ea
as representing the action $\tilde{S}_g[\tilde{g}_{\mu \nu},\phi]$ defined above. Obviously this action just propagates the two DoF of a massless spin-two field and no scalar. When disformally transformed as in (\ref{disdefinv}), however, the resulting action $S_g[g_{\mu \nu},\phi]$ contains a scalar which is not propagating either. The equivalence of the the two theories $\tilde{S}_g[\tilde{g}_{\mu \nu},\phi]$ and $S_g[g_{\mu \nu},\phi]$ has been studied in Ref.~\cite{Deruelle:2014zza}. The standard minimal matter coupling of general relativity translates into a non minimal coupling using the disformally transformed variables of the extended gravitational sector $g_{\mu \nu}$ and $\phi$. However, imagine we are just given the action $S_g[g_{\mu \nu},\phi]$ ({\it i.e.}~without knowing its equivalence with General Relativity) and couple matter minimally to the metric $g_{\mu \nu}$. A very simple such possibility is just provided by the minimal coupling of a single scalar $\Phi$ as in 
\ba
S_{m}= \int d^4 x \sqrt{-g} g^{\mu \nu}  \Phi_\mu  \Phi_\nu,
\ea
where $\Phi_\mu$ denotes $\partial_\mu \Phi$. The transformation (\ref{disdef}) then implies that this action depends explicitly on the scalar $\phi$. E.g.\ for simplicity, let us choose the functions $a$ and $b$ verifying
\ba
a&=& \sqrt{1 + \tilde{X}} \,,\\
b &=& -\frac{1}{\sqrt{1 + \tilde{X}}} 
\ea
(a choice that implies in particular $a+ b \tilde{X}=1/a$), then, it is easy to show that one has 
\ba
\tilde{S}_{m}&=& 
\int d^4 x \sqrt{-\tilde{g}}\left( \tilde{g}^{\mu \nu}   \Phi_\mu  \Phi_\nu + \phi^\mu \phi^\nu \Phi_\mu \Phi _\nu \right)
\ea
and it is clear that the theory now propagates two scalars in addition to the metric: the matter coupling has made dynamical the ``disformal" scalar. This can be verified first by writing the scalar field equations which read 
\ba
\tilde{\Box} \Phi + \tilde{\nabla}_\mu\left(\phi^\mu \Phi^\sigma \phi_\sigma\right) = 0, \\
\tilde{\nabla}_\mu\left(\Phi^\mu \phi^\sigma \Phi_\sigma\right) =0,
\ea 
where a $\tilde{}$ indicate that derivatives are computed with the $\tilde{g}_{\mu \nu}$ metric. Considering then just time dependent scalars on flat backgrounds (i.e.\ neglecting in particular the backreaction on the metric) we get, after some trivial manipulations, that these equations just boil down to the trivial (where a dot means a time derivative)
\ba
\ddot{\Phi} = 0 \,,\\
\ddot{\phi} =0
\ea
exhibiting the announced two degrees of freedom. This shows that matter cannot be consistently (in the terminology of this paper) coupled to veiled gravity. Of course veiled gravity considered as a DHOST, is somehow pathological as it is doubly degenerate: with no matter coupling, it does propagate just 2 DoF (as opposed to 3 for a more generic DHOST and 4 for a generic HOST), and this is the reason why a minimally genuine scalar $\Phi$ with minimal coupling is enough to make the disformal scalar $\varphi$ reappear as a bona fide DoF. In this case, as shown above, this just comes from the fact that the disformal transformation contains first derivatives of the scalar $\varphi$. For a generic DHOST, an increase of the DoF might stem from minimal matter coupling when the latter contains derivatives of the metric, e.g.\ connections contained in covariant derivatives, as we explore below.

\label{2.2}

\subsection{Non-canonical vector field}

Our first example is a matter vector field $B_{\mu}$ with a non-canonical kinetic term given by the action
\beq \label{eq:non-canonical vector action}
S_m=\int d^4x\sqrt{-g}\,\nabla^{\mu}B^{\nu}\nabla_{\mu}B_{\nu}\,.
\eeq
In flat spacetime this action has no constraints, and hence minimal coupling to GR trivially maintains the number of DoF, namely four. It is well known that this theory is in fact pathological on its own (see e.g.\ \cite{Riva:2005gd}), but let us emphasize again that here we are not concerned with issues such as the presence of ghost instabilities, but simply ask if the covariantized theory is consistent according to our criterion of continuity in the DoF. As a side remark, we mention that vector fields with such non-canonical kinetic terms have found applications as effective models in various contexts \cite{Landau:1986aog,Bialek:1986it,Nakayama:2010ye}.\footnote{More relevant in modified gravity is the example of Einstein--aether theory \cite{Jacobson:2000xp}, which also involves a vector field with a non-canonical kinetic Lagrangian. However this case is qualitatively different because the norm of the vector is constrained to be a constant. See also \cite{Jacobson:2011cc,Garfinkle:2012rr} for studies of the constraint structure in Einstein--aether theory. It is also interesting to remark that the infrared limit of Ho\v{r}ava gravity can be recast as a certain constrained version of the Einstein--aether model \cite{Jacobson:2010mx}.}

To settle the question we perform a Hamilton--Dirac analysis of the generic quadratic DHOST model plus the above matter action. The $3+1$ expansion of the DHOST action was given above in eq.\ \eqref{eq:3+1 full}, while for the decomposition of \eqref{eq:non-canonical vector action} we obtain
\beq\bal
S_m&=\int dtd^3x\,N\sqrt{\gamma}\Big[W_{*}^2-F^iF_i+{\cal K}^{ij,kl}_mK_{ij}K_{kl}+2B^iF^jK_{ij}\\
&\quad +2{\cal C}^{ij}_mK_{ij}-2D^iB_{*}F_i-{\cal U}_m\Big]\,.
\eal\eeq
We have redefined the $B_0$ variable as
\beq
B_{*}:=\frac{1}{N}(B_0-N^iB_i)\,,
\eeq
analogously to the DHOST vector, and introduced the shorthand notations
\beq\bal
W_{*}&:=\frac{1}{N}\left(\dot{B}_{*}-\Xi_B\right)\qquad \mbox{with}\quad \Xi_B:=B^iD_iN+N^iD_iB_{*}\,,\\
F_i&:=\frac{1}{N}\left(\dot{B}_i-\Upsilon_i\right)\qquad \mbox{with}\quad \Upsilon_i:=D_i(NB_{*})+B_kD_iN^k+N^kD_kB_i\,,\\
\eal\eeq
and the expressions
\beq\bal
{\cal K}^{ij,kl}_m&=B_{*}^2\gamma^{i(k}\gamma^{l)j}-\left(B^iB^{(k}\gamma^{l)j}+B^jB^{(k}\gamma^{l)i}\right)\,,\\
{\cal C}^{ij}_m&=2B^{(i}D^{j)}B_{*}-B_{*}D^{(i}B^{j)}\,,\\
{\cal U}_m&=2D^iB_{*}D_iB_{*}-D^iB^jD_iB_j\,.
\eal\eeq
The complete action, $S=S_g+S_m$, then reads
\beq\bal
S&=\int dtd^3x\Big\{N\sqrt{\gamma}\Big[{\cal A}V_{*}^2+2{\cal B}^{ij}V_{*}K_{ij}+{\cal K}^{ij,kl}_{\rm tot}K_{ij}K_{kl}+W_{*}^2-F^iF_i+2B^iF^jK_{ij}\\
&\quad+2{\cal C}^{ij}_{\rm tot}K_{ij}+2{\cal C}^0V_{*}-2D^iB_{*}F_i-{\cal U}_{\rm tot}\Big]\\
&\quad+\lambda^0\left(NA_{*}+N^iA_i-\dot{\phi}\right)+\lambda^i\left(A_i-D_i\phi\right)\Big\}\,,
\eal\eeq
where ${\cal K}^{ij,kl}_{\rm tot}:={\cal K}^{ij,kl}+{\cal K}^{ij,kl}_m$, ${\cal C}^{ij}_{\rm tot}:={\cal C}^{ij}+{\cal C}^{ij}_m$ and ${\cal U}_{\rm tot}:={\cal U}+{\cal U}_m$.

Proceeding with the Hamilton--Dirac analysis we first compute the canonical momenta,
\beq\begin{gathered} \label{eq:pi momenta}
\pi_0:=\frac{\partial\Lag}{\partial\dot{N}}=0\,,\qquad \pi_i:=\frac{\partial\Lag}{\partial\dot{N}^i}=0\,,\\
\pi^{ij}:=\frac{\partial\Lag}{\partial\dot{\gamma}_{ij}}=\sqrt{\gamma}\left[{\cal K}^{ij,kl}_{\rm tot}K_{kl}+{\cal B}^{ij}V_{*}+{\cal C}^{ij}_{\rm tot}+B^{(i}F^{j)}\right]\,,
\end{gathered}\eeq
\beq\begin{gathered} \label{eq:p momenta}
p_{\phi}:=\frac{\partial\Lag}{\partial\dot{\phi}}=-\lambda^0\,,\qquad p^i:=\frac{\partial\Lag}{\partial\dot{A}_i}=0\,,\\
p_{*}:=\frac{\partial\Lag}{\partial\dot{A}_{*}}=2\sqrt{\gamma}\left[{\cal A}V_{*}+{\cal B}^{ij}K_{ij}+{\cal C}^0\right]\,,
\end{gathered}\eeq
\beq \label{eq:q momenta}
q_{*}:=\frac{\partial\Lag}{\partial\dot{B}_{*}}=2\sqrt{\gamma}\,W_{*}\,,\qquad q^i:=\frac{\partial\Lag}{\partial\dot{B}_i}=2\sqrt{\gamma}\left[-F^i+B_jK^{ij}-D^iB_{*}\right]\,.
\eeq
The obvious primary constraints are again given by
\beq
\pi_0\approx0\,,\qquad \pi_i\approx0\,,\qquad p^i\approx0\,,\qquad \chi_i:=A_i-D_i\phi\approx0\,.
\eeq
The important question is whether there exists an additional constraint stemming from the degeneracy condition \eqref{eq:degeneracy cond}. It is clear however that the phase space function $\Psi$ defined in \eqref{eq:Psi function} does no longer vanish weakly. Due to the presence of the matter vector $B_{\mu}$ we instead have
\beq
\Psi\approx -2\sqrt{\gamma}\,{\cal K}^{-1}_{ij,kl}\left[{\cal K}^{ij,mn}_mK_{mn}+{\cal C}^{ij}_m+B^{(i}F^{j)}\right]{\cal B}^{kl}\,,
\eeq
which depends explicitly on the velocities and so does not define a constraint anymore. Of course it is still in principle possible that an extra constraint does exist but takes a more complicated form in the presence of matter. But in fact this is not the case, as we can demonstrate explicitly simply by showing that one can solve for the velocities $\dot{\gamma}_{ij}$, $\dot{A}_{*}$, $\dot{B}_{*}$ and $\dot{B}_i$.

From eq.\ \eqref{eq:q momenta} we immediately have $\dot{B}_{*}=\frac{Nq_{*}}{2\sqrt{\gamma}}+\Xi_B$, while from eqs.\ \eqref{eq:p momenta} and \eqref{eq:q momenta} we first find
\beq\begin{gathered} \label{eq:Adot and Bdot}
V_{*}=\frac{1}{\cal A}\left[\frac{p_{*}}{2\sqrt{\gamma}}-{\cal B}^{ij}K_{ij}-{\cal C}^0\right] \,,\\
F_i=-\frac{q_i}{2\sqrt{\gamma}}+B^jK_{ij}-D_iB_{*} \,.\\
\end{gathered}\eeq
The latter can be substituted in \eqref{eq:pi momenta} to get the following equation for $K_{ij}$:
\beq\bal
&\left({\cal K}^{ij,kl}-\frac{{\cal B}^{ij}{\cal B}^{kl}}{{\cal A}}\right)K_{kl}+B_{*}^2K^{ij}-B_kB^{(i}K^{j)k}\\
&=\frac{1}{\sqrt{\gamma}}\left(\pi^{ij}+\frac{B^{(i}q^{j)}}{2}\right)+B^{(i}D^{j)}B_{*}-\frac{1}{\cal A}\left(\frac{p_{*}}{2\sqrt{\gamma}}-{\cal C}^0\right){\cal B}^{ij}-{\cal C}^{ij}_{\rm tot}\,,
\eal\eeq
and observe that the RHS depends only on the canonical variables. Although the matrix ${\cal K}^{ij,kl}-\frac{{\cal B}^{ij}{\cal B}^{kl}}{{\cal A}}$ is non-invertible by virtue of the degeneracy condition, eq.\ \eqref{eq:degeneracy cond}, the presence of the field $B_{\mu}$ renders this equation invertible (for generic field values), and therefore the metric velocity $\dot{\gamma}_{ij}$ can be solved for in terms of the canonical variables. The result can then be substituted back into eqs.\ \eqref{eq:Adot and Bdot} to determine $\dot{A}_{*}$ and $\dot{B}_i$.

The analysis of secondary constraints proceeds almost identically to the vacuum case of sec.\ \ref{sec:dhost}, with the simplification that the DHOST constraint is now absent. The augmented Hamiltonian thus takes the form
\beq
H_{\rm aug}=\int d^3x\left[N{\cal H}_0+N^i{\cal H}_i+\mu^0\pi_0+\mu^i\pi_i+\lambda^i\chi_i+\alpha_ip^i\right]\,.
\eeq
Although the Hamiltonian and momentum constraints, ${\cal H}_0\approx0$ and ${\cal H}_i\approx0$, now receive contributions from the matter vector field, general covariance again guarantees that they will be first class. Moreover, the momentum $p_{\phi}$ appears in ${\cal H}_0$ and ${\cal H}_i$ in the same way as in the vacuum case: ${\cal H}_0\supset p_{\phi}A_{*}$ and ${\cal H}_i\supset p_{\phi}A_i$. The preservation in time of the constraint $\chi_i\approx0$ therefore determines the Lagrange multiplier $\alpha_i$ through exactly the same relation obtained above in \eqref{eq:lagrange multiplier sol}. Similarly the consistency of the constraint $p^i\approx0$ simply yields $\lambda^i=0$ by the arguments already given.

The final tally of DoF is as follows. The fields $(g_{\mu\nu},A_{\mu},\phi,B_{\mu})$ span a $38=19\times2$ dimensional phase space. Subtracting $2\times 8$ for the first class constraints and $6$ for the second class constraints, we get 16, that is 8 DoF in total. Since the vector $B_{\mu}$ propagates 4 DoF, this means that the gravitational sector has 4 DoF, which is 1 more than in vacuum. The conclusion is that the coupling to the matter vector has spoiled the degeneracy of the gravitational action with the result that the Ostrogradski ghost has reappeared.

\subsection{Cubic galileon}

The second example of matter field that we study is the cubic galileon,
\beq
S_m=\int d^4x\sqrt{-g}\bigg[-\frac{1}{2}(\nabla\pi)^2+\kappa(\nabla\pi)^2\Box\pi\bigg]\,,
\eeq
with $\kappa$ a constant. For the Hamilton--Dirac analysis we write the action with only first derivatives by introducing an auxiliary vector $B_{\mu}$ constrained as $B_{\mu}=\nabla_{\mu}\pi$,
\beq
S_m=\int d^4x\bigg\{\sqrt{-g}\bigg[-\frac{1}{2}\,B^{\mu}B_{\mu}+\kappa B^{\mu}B_{\mu}\nabla_{\nu}B^{\nu}\bigg]+\sigma^{\mu}(B_{\mu}-\nabla_{\mu}\pi)\bigg\}\,,
\eeq
and $\sigma^{\mu}$ is a Lagrange multiplier.

The $3+1$ decomposition is straightforward and in fact very analogous to that of the DHOST action, since in fact the cubic galileon falls in the same class. We find
\beq\bal
S_m&=\int dtd^3x\Big\{N\sqrt{\gamma}\Big[2{\cal C}^{ij}_mK_{ij}+\kappa(B_{*}^2-B^2)W_{*}-{\cal U}_m\Big]\\
&\quad+\sigma^0\left(NB_{*}+N^iB_i-\dot{\pi}\right)+\sigma^i\left(B_i-D_i\pi\right)\Big\}\,,
\eal\eeq
where
\beq\bal
{\cal C}^{ij}_m&=\frac{\kappa}{2}\,B_{*}(B_{*}^2-B^2)\gamma^{ij}\,,\\
{\cal U}_m&=-\frac{1}{2}(B_{*}^2-B^2)+\kappa(B_{*}^2-B^2)D_iB^i\,,
\eal\eeq
and $B^2:= B^iB_i$.

The complete action $S=S_g+S_m$ is then given by
\beq\bal
S&=\int dtd^3x\Big\{N\sqrt{\gamma}\Big[{\cal A}V_{*}^2+2{\cal B}^{ij}V_{*}K_{ij}+{\cal K}^{ij,kl}K_{ij}K_{kl}\\
&\quad+2{\cal C}^{ij}_{\rm tot}K_{ij}+2{\cal C}^0V_{*}+\kappa(B_{*}^2-B^2)W_{*}-{\cal U}_{\rm tot}\Big]\\
&\quad+\lambda^0\left(NA_{*}+N^iA_i-\dot{\phi}\right)+\lambda^i\left(A_i-D_i\phi\right)\\
&\quad+\sigma^0\left(NB_{*}+N^iB_i-\dot{\pi}\right)+\sigma^i\left(B_i-D_i\pi\right)\Big\}\,,
\eal\eeq
with ${\cal C}^{ij}_{\rm tot}:={\cal C}^{ij}+{\cal C}^{ij}_m$ and ${\cal U}_{\rm tot}:={\cal U}+{\cal U}_m$. Observe that, unlike in the previous example, the kinetic terms do not receive contributions from the matter field. We thus expect that the primary constraints of DHOST theory to remain, although we will see that this is not enough to guarantee the consistency of the model in the presence of matter.

To settle this we proceed with the Hamilton--Dirac analysis, starting with the canonical momenta,
\beq\begin{gathered}
\pi_0:=\frac{\partial\Lag}{\partial\dot{N}}=0\,,\qquad \pi_i:=\frac{\partial\Lag}{\partial\dot{N}^i}=0\,,\\
\pi^{ij}:=\frac{\partial\Lag}{\partial\dot{\gamma}_{ij}}=\sqrt{\gamma}\left[{\cal K}^{ij,kl}K_{kl}+{\cal B}^{ij}V_{*}+{\cal C}^{ij}_{\rm tot}\right]\,,
\end{gathered}\eeq
\beq\begin{gathered}
p_{\phi}:=\frac{\partial\Lag}{\partial\dot{\phi}}=-\lambda^0\,,\qquad p^i:=\frac{\partial\Lag}{\partial\dot{A}_i}=0\,,\\
p_{*}:=\frac{\partial\Lag}{\partial\dot{A}_{*}}=2\sqrt{\gamma}\left[{\cal A}V_{*}+{\cal B}^{ij}K_{ij}+{\cal C}^0\right]\,,
\end{gathered}\eeq
\beq\begin{gathered}
q_{\pi}:=\frac{\partial\Lag}{\partial\dot{\pi}}=-\sigma^0\,,\qquad q^i:=\frac{\partial\Lag}{\partial\dot{B}_i}=0\,,\\
q_{*}:=\frac{\partial\Lag}{\partial\dot{B}_{*}}=\sqrt{\gamma}\,\kappa(B_{*}^2-B^2)\,.
\end{gathered}\eeq
We can immediately read off the primary constraints
\beq\begin{gathered}
\pi_0\approx0\,,\qquad \pi_i\approx0\,,\qquad p^i\approx0\,,\qquad q^i\approx0\,,\\
\chi_i:=A_i-D_i\phi\approx0\,,\qquad \psi_i:=B_i-D_i\pi\approx0\,,
\end{gathered}\eeq
in addition to the two constraints associated to the degeneracies of the action,
\beq\bal
\Psi'&:= p_{*}-2{\cal K}^{-1}_{ij,kl}\pi^{ij}{\cal B}^{kl}+2\sqrt{\gamma}\left({\cal K}^{-1}_{ij,kl}{\cal C}^{ij}_{\rm tot}{\cal B}^{kl}-{\cal C}^0\right)\approx0\,,\\
\Lambda&:= q_{*}-\sqrt{\gamma}\,\kappa(B_{*}^2-B^2)\approx0\,.
\eal\eeq
Note that $\Psi'$ differs from the vacuum constraint $\Psi$ in \eqref{eq:Psi function} in that it involves the vector $B_{\mu}$ contained in ${\cal C}^{ij}_{\rm tot}$.

The augmented Hamiltonian then reads
\beq
H_{\rm aug}=\int d^3x\left[N{\cal H}_0+N^i{\cal H}_i+\mu^0\pi_0+\mu^i\pi_i+\lambda^i\chi_i+\alpha_ip^i+\sigma^i\psi_i+\beta_iq^i+\xi\Psi'+\rho\Lambda\right]\,,
\eeq
where $\mu^0$, $\mu^i$, $\lambda^i$, $\alpha_i$, $\sigma^i$, $\beta_i$, $\xi$ and $\rho$ form the set of Lagrange multipliers at this stage in the analysis, and the Hamiltonian and momentum constraint functions are given explicitly by
\beq\bal
{\cal H}_0&=\sqrt{\gamma}\left[{\cal K}^{-1}_{ij,kl}\left(\frac{\pi^{ij}}{\sqrt{\gamma}}-{\cal C}^{ij}_{\rm tot}\right)\left(\frac{\pi^{kl}}{\sqrt{\gamma}}-{\cal C}^{kl}_{\rm tot}\right)+{\cal U}_{\rm tot}\right]\\
&\quad +p_{\phi}A_{*}-D_i(p_{*}A^i)+q_{\pi}B_{*}-D_i(q_{*}B^i)\,,\\
{\cal H}_i&=-2D^j\pi_{ij}+p_{\phi}A_i+p_{*}D_iA_{*}+q_{\pi}B_i+q_{*}D_iB_{*}\,.
\eal\eeq

Continuing with the analysis, we immediately obtain the diffeomorphism secondary constraints ${\cal H}_0\approx0$ and ${\cal H}_i\approx0$, while the consistency of the constraints $p^i\approx0$, $\chi_i\approx0$, $q^i\approx0$ and $\psi_i\approx0$ determines the associated Lagrange multipliers in the by now familiar way:
\beq\bal
0&=\{\chi_i,H_{\rm aug}\}=\alpha_i+D_i(NA_{*}+N^jA_j)\qquad \Rightarrow\qquad \alpha_i=-D_i(NA_{*}+N^jA_j)\,,\\
0&=\{p^i,H_{\rm aug}\}=-\lambda^i\qquad \Rightarrow\qquad \lambda^i=0\,,\\
0&=\{\psi_i,H_{\rm aug}\}=\beta_i+D_i(NB_{*}+N^jB_j)\qquad \Rightarrow\qquad \beta_i=-D_i(NB_{*}+N^jB_j)\,,\\
0&=\{q^i,H_{\rm aug}\}=-\sigma^i\qquad \Rightarrow\qquad \sigma^i=0\,.
\eal\eeq
We have seen that in the absence of matter the constraint $\Psi\approx0$ led to a secondary constraint, whose own consistency would then determine the multiplier $\xi$. With galileon matter the story is modified because of the fact that $\Psi'$ and $\Lambda$ do not commute. Explicitly,\footnote{Here we are omitting the coordinates to avoid cluttering. The full expression should of course read $\left\{\Psi'(t,\mathbf{x}),\Lambda(t,\mathbf{y})\right\}=(\ldots)\delta(\mathbf{x}-\mathbf{y})$.}
\beq
\left\{\Psi',\Lambda\right\}=2\kappa\sqrt{\gamma}\,{\cal K}^{-1}_{ij,kl}\left(B_{*}^2\gamma^{ij}-B^iB^j\right){\cal B}^{kl}\,.
\eeq
Therefore the preservation in time of $\Psi'\approx0$ and $\Lambda\approx0$ will determine the Lagrange multipliers $\xi$ and $\rho$ instead of producing any secondary constraints, and the Hamilton--Dirac analysis ends at this point.

To summarize, there are 8 first class constraints associated to general covariance while the rest, 14 in total, are all second class constraints. The fields $(g_{\mu\nu},A_{\mu},\phi,B_{\mu},\pi)$ span a $40=20\times2$ dimensional phase space, from which we subtract $2\times8+14$ to get $10$, that is 5 DoF. This is again 1 too many for a healthy scalar-tensor theory coupled to a galileon matter field, from which we infer the presence of an Ostrogradski ghost.


\section{Spinor fields in DHOST} \label{sec:spinors}

In this section we consider the minimal coupling of a spinor field to DHOST. Spinors fall in the class of ``dangerous'' matter fields for the simple reason that their kinetic Lagrangian includes the spin connection, and so time derivatives of the tetrad. Moreover their equations of motion are first order and they feature constraints. We will however demonstrate that the coupling of spinors to DHOST is in fact consistent in that all constraints are preserved.\footnote{See \cite{Gauthier:2009wc} for earlier work on spinors in (non-degenerate) scalar-tensor theories of gravity. Also related to the context of DHOST theories are the analyses of \cite{Domenech:2015hka,Bittencourt:2015ypa} who considered the behavior of a minimally coupled spinor under disformal transformations, and \cite{Kimura:2017gcy,Kimura:2018sfs} who studied degenerate scalar-spinor systems without gravity.} We will focus on the treatment of a massless spin-1/2 Majorana spinor for the sake of simplicity (a Dirac spinor would pose no further trouble than doubling the spinor's DoF); more involved cases such as the analysis of a spin-3/2 field would be very interesting but certainly beyond our present scope.

\subsection{Majorana field in DHOST}

The minimally covariantized action of a massless spin-1/2 Majorana spinor $\lambda_{\alpha}$ is given by\footnote{We employ the spinor conventions of \cite{Freedman:2012zz}. In particular, we use 4-component notation for our spinor field $\lambda_{\alpha}$, with the Majorana conjugate being denoted by $\lambda^{\alpha}:=C^{\alpha\beta}\lambda_{\beta}$, where the matrix $C^{\alpha\beta}$ is related to the charge conjugation matrix. Our gamma matrices are always the standard constant ones of flat spacetime (we will {\it not} use the notation $\gamma^{\mu}=e_a^{\phantom{a}\mu}\gamma^a$ in order to avoid any confusion), and a multi-index gamma matrix stands for the anti-symmetrized product: $\gamma^{a_1\cdots a_n}:=\gamma^{[a_1}\cdots\gamma^{a_n]}$. Lorentz indices are raised and lowered with the Minkowski metric $\eta_{ab}$ (we will refrain from moving the spacetime index in the tetrads, again to avoid confusion). \label{fn:spinor conventions}}
\beq
S_m=-\frac{1}{2}\int d^4x\sqrt{-g}\,e_a^{\phantom{a}\mu}\lambda^{\alpha}(\gamma^a)_{\alpha}^{\phantom{\alpha}\beta}\nabla_{\mu}\lambda_{\beta}\,,
\eeq
where $e_a^{\phantom{a}\mu}$ is the inverse of the tetrad field $e^a_{\phantom{a}\mu}$, related to the metric via $g_{\mu\nu}=\eta_{ab}e^a_{\phantom{a}\mu}e^b_{\phantom{b}\nu}$, while the covariant derivative of $\lambda_{\alpha}$ reads
\beq
\nabla_{\mu}\lambda_{\alpha}=\partial_{\mu}\lambda_{\alpha}+\frac{1}{4}\,\omega^{ab}_{\phantom{ab}\mu}(\gamma_{ab})_{\alpha}^{\phantom{\alpha}\beta}\lambda_{\beta}\,,
\eeq
and $\omega^{ab}_{\phantom{ab}\mu}$ is the spin connection. In the second order formalism that we adopt it can be expressed as $\omega^a_{\phantom{a}b\mu}=e_b^{\phantom{b}\nu}(\Gamma^{\rho}_{\mu\nu}e^a_{\phantom{a}\rho}-\partial_{\mu}e^a_{\phantom{a}\nu})$.

Here and in the following subsection, we will consider, as is customary (see e.g.\ \cite{DeWitt:1992cy,Henneaux:1992ig,Prokhorov:2011zz}), all (classical) variables and fields as supernumber-valued. {\it I.e.} our variables will take their values in an infinite dimensional Grassmann algebra with the infinite set of (anticommuting) generators $\theta^A$ such that any supernumber $z$ can be written as (with the Einstein summation implied on the indices $A,B$ running over the generators)
\ba
z = z^{(0)} + z_{A}\theta^A+z_{AB}\theta^A\theta^B+\ldots 
\ea 
where $z^{(0)}$, $z_{A}$, $z_{AB}$, $\dots$ are just real or complex numbers. All terms with the same number of generators on the right hand side of the above equation, say $n$, belong to what is called the level $n$ component of the supernumber. The level 0 component $z^{(0)}$ is called the body of the supernumber, while the other level $n$, with $n \geq 1$, sum up to what is called the soul $z_{S}$ of $z$. Hence we can write 
\ba
z = z^{(0)} + z_{S}\,. 
\ea
For future reference we can also define as $z^{(n)}$ the sum of all components at level $n$, e.g.\
\ba
z^{(1)} = z_{A}\theta^A, \quad z^{(2)} = z_{AB}\theta^A\theta^B, \quad \ldots
\ea
such that $z_{S} = \sum_{n=1}^{n= \infty} z^{(n)}$. Fermionic variables only have non vanishing odd level components (and hence in particular a vanishing body), while bosonic variables only have non vanishing even level components. Note that, as usual, for consistency, bosonic variables do not only have a non-trivial body, but should also be considered as having a non-trivial soul.

The Hamilton--Dirac analysis of GR in the tetrad formalism and with spinorial matter has been studied in \cite{Deser:1976ay,Nelson:1977qj,Nelson:1978ex,Henneaux:1978wlm,Charap:1988vz}, and most of the results carry over to DHOST gravity in a rather straightforward way.
Following \cite{Henneaux:1978wlm} we first introduce the matrices
\beq
E^0:=-n^{\mu}e^a_{\phantom{a}\mu}\gamma_a\,,\qquad E^i:=\gamma^{ij}e^a_{\phantom{a}j}\gamma_a\,.
\eeq
Note that in terms of ADM variables we have $E^0=Ne_a^{\phantom{a}0}\gamma^a$, which can be regarded as a redefinition of the variable $e_a^{\phantom{a}0}$ that will have the effect of removing non-linear terms in the lapse (very analogously to the redefinition of $A_0$ in eq.\ \eqref{eq:Astar def}).\footnote{The explicit inverse relation that expresses $e_a^{\phantom{a}0}$ in terms of $E^0$ is $e_a^{\phantom{a}0}=\frac{1}{2N}\,\{E^0,\gamma_a\}$.} The canonical momentum of the spinor field can then be written as\footnote{Here $\partial^R$ stands for right-differentiation; see e.g.\ \cite{Henneaux:1992ig,Prokhorov:2011zz} for general discussions on the canonical formalism with Grassmann-odd variables.}
\beq
\varpi^\alpha:=\frac{\partial^R\Lag}{\partial\dot{\lambda}_{\alpha}}=-\frac{1}{2}\,\sqrt{\gamma}\,\lambda^{\beta}(E^0)_{\beta}^{\phantom{\beta}\alpha}\,.
\eeq
As remarked at the beginning of the section, the gravitational coupling of the spinor also modifies the graviton's canonical momentum. A direct calculation yields
\beq\begin{gathered}
\pi_a^{\phantom{a}0}:=\frac{\partial\Lag}{\partial\dot{e}^a_{\phantom{a}0}}=0 \,,\\
\pi_a^{\phantom{a}i}:=\frac{\partial\Lag}{\partial\dot{e}^a_{\phantom{a}i}}=2\sqrt{\gamma}\,e_{aj}\left[{\cal K}^{ij,kl}K_{kl}+{\cal B}^{ij}V_{*}+{\cal C}^{ij}\right]+\frac{1}{8}\,\sqrt{\gamma}\,\lambda^{\alpha}(\gamma_aE^iE^0)_{\alpha}^{\phantom{\alpha}\beta}\lambda_{\beta} \,.\\
\end{gathered}\eeq
Notice that the momentum corresponding to the 3-metric (which in the tetrad formalism should be taken merely as a shorthand notation since $\gamma_{ij}$ is not an independent variable) does not receive contributions from the spinor field,\footnote{To see this step more explicitly, observe that the matter contribution to $\pi^{ij}$ is $\frac{1}{16}\,\sqrt{\gamma}\,\lambda^{\alpha}(E^{(i}E^{j)}E^0-\frac{1}{N}E^0N^{(i}E^{j)}E^0)_{\alpha}^{\phantom{\alpha}\beta}\lambda_{\beta}$, which follows from the identity $e_a^{\phantom{a}i}\gamma^a=E^i-\frac{N^i}{N}\,E^0$. On using the gamma matrix Clifford algebra the previous expression reduces to a sum of spinor bilinears of the form $\lambda^{\alpha}(\gamma_a)_{\alpha}^{\phantom{\alpha}\beta}\lambda_{\beta}$, which vanish due to a Majorana flip identity.}
\beq \label{eq:pi momentum spinor}
\pi^{ij}:=\frac{1}{2}\,e^{a(i}\pi_a^{\phantom{a}j)}=\sqrt{\gamma}\left[{\cal K}^{ij,kl}K_{kl}+{\cal B}^{ij}V_{*}+{\cal C}^{ij}\right]\,,
\eeq
and of course coincides with the result of vacuum DHOST (a property that is well known in GR \cite{Henneaux:1978wlm}). For completeness we also write again the momenta associated to the DHOST scalar and auxiliary vector, which are trivially the same as in vacuum:
\beq\begin{gathered}
p_{\phi}:=\frac{\partial\Lag}{\partial\dot{\phi}}=-\lambda^0\,,\qquad p^i:=\frac{\partial\Lag}{\partial\dot{A}_i}=0\,,\\
p_{*}:=\frac{\partial\Lag}{\partial\dot{A}_{*}}=2\sqrt{\gamma}\left[{\cal A}V_{*}+{\cal B}^{ij}K_{ij}+{\cal C}^0\right]\,.
\end{gathered}\eeq

From these results we next infer the set of primary constraints. For the tetrad sector we have\footnote{In deriving the local Lorentz constraint it proves useful to know the identity $\gamma^{ij}e^a_{\phantom{a}i}e^b_{\phantom{a}j}=\eta^{ab}+n^{\mu}n^{\nu}e^a_{\phantom{a}\mu}e^b_{\phantom{a}\nu}$.}
\beq
\pi_a^{\phantom{a}0}\approx0\,,\qquad J_{ab}:=\pi_{[a}^{\phantom{[a}i}e_{b]i}+\frac{1}{4}\,\varpi^{\alpha}(\gamma_{ab})_{\alpha}^{\phantom{\alpha}\beta}\lambda_{\beta}\approx0\,,
\eeq
and correspond to the primary constraints associated to general covariance and local Lorentz invariance, respectively \cite{Deser:1976ay}. For the spinor sector we have the standard constraint
\beq \label{eq:spinor constraint}
\Lambda^{\alpha}:=\varpi^\alpha+\frac{1}{2}\,\sqrt{\gamma}\,\lambda^{\beta}(E^0)_{\beta}^{\phantom{\beta}\alpha}\approx0\,,
\eeq
from the fact that the Lagrangian is linear in the spinor's velocity. Lastly we have the familiar DHOST constraints
\beq\begin{gathered}
p^i\approx0\,,\qquad \chi_i:=A_i-D_i\phi\approx0 \,,\\
\Psi:=p_{*}-2{\cal K}^{-1}_{ij,kl}\pi^{ij}{\cal B}^{kl}+2\sqrt{\gamma}\left({\cal K}^{-1}_{ij,kl}{\cal C}^{ij}{\cal B}^{kl}-{\cal C}^0\right)\approx 0\,,
\end{gathered}\eeq
the latter being a direct consequence of eq.\ \eqref{eq:pi momentum spinor} and the degeneracy condition \eqref{eq:degeneracy cond}.

After some further manipulations we derive the augmented Hamiltonian,
\beq
H_{\rm aug}=\int d^3x\left[N{\cal H}_0+N^i{\cal H}_i+\mu^a\pi_a^{\phantom{a}0}+\epsilon^{ab}J_{ab}+\lambda^i\chi_i+\alpha_ip^i+\xi\Psi+\Lambda^{\alpha}\zeta_{\alpha}\right]\,,
\eeq
where
\beq\bal
{\cal H}_0&=\sqrt{\gamma}\left[{\cal K}^{-1}_{ij,kl}\left(\frac{\pi^{ij}}{\sqrt{\gamma}}-{\cal C}^{ij}\right)\left(\frac{\pi^{kl}}{\sqrt{\gamma}}-{\cal C}^{kl}\right)+{\cal U}\right]+p_{\phi}A_{*}-D_i(p_{*}A^i)\\
&\quad +\frac{1}{2}\,\sqrt{\gamma}\,\lambda^{\alpha}(E^i)_{\alpha}^{\phantom{\alpha}\beta}D_i\lambda_{\beta}\,,\\
{\cal H}_i&=-2D^j\pi_{ij}+p_{\phi}A_i+p_{*}D_iA_{*}\\
&\quad -\frac{1}{2}\,\sqrt{\gamma}\,\lambda^{\alpha}(E^0)_{\alpha}^{\phantom{\alpha}\beta}D_i\lambda_{\beta}+\frac{1}{8}\,\sqrt{\gamma}\,D_j\left[\lambda^{\alpha}(E^jE_iE^0)_{\alpha}^{\phantom{\alpha}\beta}\lambda_{\beta}\right] \,,
\eal\eeq
and $\mu^a$, $\epsilon^{ab}$, $\lambda^i$, $\alpha_i$, $\xi$ and $\zeta_{\alpha}$ are Lagrange multipliers. The analysis of secondary constraints proceeds much like in the examples of sec.\ \ref{sec:matter}. Once again the general covariance of the action guarantees that the constraints $\pi_a^{\phantom{a}0}\approx0$ and their descendants ${\cal H}_0\approx0$ and ${\cal H}_i\approx0$ are first class. By the same reasoning, the local Lorentz invariance of the action in tetrad variables ensures that the $J_{ab}\approx0$ are also first class (but they do not generate secondary constraints \cite{Nelson:1977qj}). It is also obvious that the preservation in time of the constraints $\chi_i\approx0$ and $p^i\approx0$ works out exactly as in vacuum, since the contributions of the matter spinor are unrelated to $\phi$ and $A_{\mu}$, with the familiar result that they simply determine the multipliers $\lambda^i$ and $\alpha_i$.

The only outstanding question is whether constraints $\Psi\approx0$ and $\Lambda^{\alpha}\approx0$ give rise to secondary constraints. Recall that this was precisely the step where the galileon matter field evinced its inconsistency, which we saw to stem from the fact that the primary constraints did not commute. The case at hand is however slightly more complex as the relevant question is whether the matrix
\beq
{\cal M}^{IJ}:=\left\{{\cal C}^I,{\cal C}^J\right\}\,,\qquad \mbox{with}\quad {\cal C}^I:=\left(\Psi,\Lambda^{\alpha}\right)\,,
\eeq
possesses an inverse or not. More precisely, the rank of ${\cal M}$ equals the number of Lagrange multipliers among the set $(\xi,\zeta_{\alpha})$ that are fixed by the requirement of preservation in time of the primary constraints.

The matrix ${\cal M}$ is in fact singular. Working out the Poisson brackets we find\footnote{Here we raise a spinor index in the matrix $E^0$ using the charge conjugation matrix (see footnote \ref{fn:spinor conventions}). Note that $(E^0)^{\alpha\beta}$ is symmetric, in accordance with the fact that the Poisson bracket of two Grassmann-odd functions must be symmetric.}
\beq
\{\Psi,\Lambda^{\alpha}\}=\frac{1}{2}\,\sqrt{\gamma}\,{\cal K}^{-1}_{ij,kl}\gamma^{ij}{\cal B}^{kl}\lambda^{\beta}(E^0)_{\beta}^{\phantom{\beta}\alpha}\,,\qquad \{\Lambda^{\alpha},\Lambda^{\beta}\}=\sqrt{\gamma}(E^0)^{\alpha\beta}\,.
\eeq
Deciding whether ${\cal M}$ is invertible turns out to be very easy and follows directly from a simple lemma on the invertibility of Grassmann-valued matrices \cite{DeWitt:1992cy,Henneaux:1992ig}: given the decomposition of any matrix ${\cal M}$ into its body and soul, ${\cal M}={\cal M}^{(0)}+{\cal M}_S$, then ${\cal M}$ is invertible if and only if the body ${\cal M}^{(0)}$ is invertible. In our case we have
\beq
{\cal M}^{(0)}=\left(\begin{array}{cc}
0 & 0 \\
0 & \sqrt{\gamma}(E^0)^{\alpha\beta}
\end{array}\right)\,,
\eeq
which is obviously singular. Moreover, since the matrix $\{\Lambda^{\alpha},\Lambda^{\beta}\}$ is non-singular (both the charge conjugation matrix and the gamma matrices are non-singular), we infer that ${\cal M}$ has rank four and therefore the time preservation of the constraints $\Psi\approx0$ and $\Lambda^{\alpha}\approx0$ leads to precisely one secondary constraint (the constraint $\Omega\approx0$ explained in sec.\ \ref{sec:dhost}) and the determination of four Lagrange multipliers.

The merry conclusion is that the DHOST constraint remains in the presence of spinorial matter and the Ostrogradski ghost is absent. Indeed we can complete the counting of DoF by taking the $25\times2$ phase space variables\footnote{Although the four components in the spinor $\lambda_{\alpha}$ could be complex, the Majorana condition implies that there are four independent real variables. Note that in four dimensions one may use a so-called ``really real'' representation of the gamma matrices, in which case Majorana spinors are purely real \cite{Freedman:2012zz}.} corresponding to the fields $(e^a_{\phantom{a}\mu},A_{\mu},\phi,\lambda_{\alpha})$, subtract $14\times2$ for the first class constraints $\{\pi_a^{\phantom{a}0},{\cal H}_{\mu},J_{ab}\}\approx0$ and $12$ for the second class constraints $\{p^i,\chi_i,\Psi,\Omega,\Lambda^{\alpha}\}\approx0$, giving a grand total of $10$, that is $5$ DoF, which matches the expectation of having $3$ DoF for the scalar-tensor sector plus $2$ DoF for the Majorana spinor field.

One may worry that this outcome may be specific to the simple spinor action that we have considered, which neglects all self-interactions. It is obvious however that any non-derivative potential will have no effect on the constraint structure and hence is perfectly allowed. At the next level of complexity one may consider arbitrary interactions that are linear in the derivative of the spinor field. The latter assumption is sufficient in order to maintain an analogue of the constraint \eqref{eq:spinor constraint}. What is less trivial is that this condition also ensures that, upon minimal coupling to gravity, the 3-metric canonical momentum defined in \eqref{eq:pi momentum spinor} remains independent of the spinor field (the proof of this property is somewhat technical and is given in appendix \ref{sec:appendix}), and this is enough to guarantee that the DHOST constraint is left unchanged. It thus follows that any self-interacting spin-1/2 particle admits a consistent coupling to DHOST provided the matter action is at most linear in the spinor derivative. It is worth emphasizing that this requirement is not very restrictive, since generalized spinor models that involve second or higher derivatives or terms non-linear in first derivatives generically lead to the loss of the constraint \eqref{eq:spinor constraint}, or equivalently to higher than first order eqs.\ of motion, which typically signals a pathology. It is only very recently that healthy generalized fermionic systems have been constructed \cite{Kimura:2018sfs}, although the consistency of the coupling to gravity for such models, not only in DHOST but already in pure GR, remains an open question that we would like to address in the future.

\subsection{Superclassical dynamics with higher derivatives}

It is evident from the previous analysis that the consistent coupling between DHOST and a spinor matter field has little to do with the detailed structure of the scalar-tensor sector or with the specific form of the constraints. It is rather due to the properties of the spinor fields and the assumption that their action is at most linear in time derivatives, which we have seen to be a sufficient condition for the existence of the spinor and DHOST primary constraints in the minimally coupled system. The fact that a secondary scalar constraint is also guaranteed to exist is then immediate from the properties of classical fermionic fields, and particularly from the lemma quoted in the previous subsection. This suggests that the consistency of interactions between higher-derivative bosonic systems and a fermionic sector is very generic, thus potentially opening the door to a wealth of novel theories.

Here we would like to initiate an exploration of this question by considering a mechanical toy model involving a set of commuting and anti-commuting variables, i.e.\ a superclassical mechanical model (see e.g.\ \cite{DeWitt:1992cy}). Concretely we envisage the action
\beq \label{eq:toy model action}
S=\int dt\left[\phi\dot{\pi}+\frac{1}{2}\,\chi\dot{\chi}+\frac{1}{2}\,\xi\dot{\xi}+g\chi\xi\phi\ddot{\pi}\right]\,,
\eeq
where $\phi$ and $\pi$ are commuting and $\chi$ and $\xi$ are anti-commuting supernumber-valued time dependent functions, while $g$ is a coupling constant. Morally we can think of the $(\phi,\pi)$ subsystem as a toy scalar-tensor theory, while $(\chi,\xi)$ represents some fermionic matter sector. We are simplifying the discussion as much as possible by considering a situation in which, in the absence of coupling ($g=0$), the dynamics is purely of first differential order. It is thus evident that the number of DoF (in the mechanical sense) is four when the two subsectors do not interact. The question is whether this conclusion changes by the inclusion of the above quartic interaction, which modifies {\it a priori} the differential order of the eqs.\ of motion. Note that the above interaction is chosen in close analogy with the fermionic coupling to gravity of the previous subsection: there, indeed, we stressed that the spin connection introduces a coupling to the fermionic fields which is (i) quadratic in the fermionic fields and (ii) depending linearly on the first derivative of the bosonic tetrad which in turn combines into the bosonic metric. Carrying then a disformal transformation, we can expect to generate in this way a coupling between the second derivative of the scalar of the disformal transformation and a term quadratic in the fermions, just like the interaction term above, while for DHOST theories obtained by disformally transforming a Horndeski theory the pure bosonic sector has just second order field equations (after the disformal transformation).

After some simple manipulations of the Euler--Lagrange equations we arrive at the following differential system:
\beq\begin{gathered} \label{eq:toy model eom}
\dot{\pi}=-g\chi\xi\ddot{\pi}\,,\qquad \dot{\phi}=g\chi\xi\ddot{\phi}\,,\\
\dot{\chi}=-g\xi\phi\ddot{\pi}\,,\qquad \dot{\xi}=-g\chi\phi\ddot{\pi}\,,
\end{gathered}
\eeq
which is indeed higher order. However, the variables being here supervariables, this implies that this system demultiplies into a recursive system at each level of the Grassmann algebra. This should be taken into account in order to properly count the number of DoF (see e.g.\ \cite{ChoquetBruhat:1984hn} where a similar analysis is carried out to address the Cauchy problem of supergravity). More specifically, we write here 
\beq
\begin{gathered}
\pi=\pi^{(0)}(t)+\pi^{(2)}(t) + \ldots\, \qquad    \chi=\chi^{(1)}(t) + \chi^{(3)}(t) + \ldots \,
\\
\phi=\phi^{(0)}(t)+\phi^{(2)}(t) + \ldots\,\qquad  \xi=\xi^{(1)}(t) + \xi^{(3)}(t) +\ldots\,
\end{gathered}
\eeq
At level 0 in the Grassmann algebra the equations \eqref{eq:toy model eom} give the free system 
\beq\begin{gathered} \label{eq:toy model eom level 0}
\dot{\pi}^{(0)}=0\,,\qquad \dot{\phi}^{(0)}=0\,
\end{gathered}
\eeq with trivial solution $\pi^{(0)}(t)=c_\pi$ and $\phi^{(0)}(t)=c_\phi$, $c_\pi$ and $c_\phi$ being c-numbers. At level 1 the system \eqref{eq:toy model eom} boils down to 
\beq\begin{gathered} \label{eq:toy model eom level 1}
\dot{\chi}^{(1)}=-g\xi^{(1)}\phi^{(0)}\ddot{\pi}^{(0)}\,,\qquad \dot{\xi}^{(1)}=-g\chi^{(1)}\phi^{(0)}\ddot{\pi}^{(0)}\,
\end{gathered}
\eeq
where the right hand sides of the above equations vanish by virtue of the body level system \eqref{eq:toy model eom level 0}. Hence, the fermionic level 1 is again free and only needs two integration constants. 
At level 2 we have 
\beq\begin{gathered}
\dot{\pi}^{(2)}=-g\chi^{(1)}\xi^{(1)}\ddot{\pi}^{(0)}\,,\qquad \dot{\phi}^{(2)}=g\chi^{(1)}\xi^{(1)}\ddot{\phi}^{(0)}
\end{gathered}
\eeq
where we see that the quartic interaction does not introduce more integration constants than in the non-interacting case which would yield the above equations with vanishing right hand sides. Indeed, at level 2, the interaction only appears on the right hand side of the above and involves only components of levels strictly smaller than 2 which have been previously determined (moreover, at level 2, the right hand sides above in fact vanish as a consequence of the body equations). It is easy to see that this statement persists at all levels. This settles the question about the number of DoF: the non-linear coupling has no effect once the level decomposition of the variables is performed and we can conclude that the interacting theory has the same number of DoF as the free one. Somehow, the Grassmann algebra level decomposition forces one to consider the interaction in a perturbative way and does not allow it to introduce extra degrees of freedom. This is also the essence of the result of the previous subsection.

The previous analysis was of course almost trivial due to the simplicity of the model and the fact that the eqs.\ of motion could be recast in a form that made manifest the structure of the level decomposition. For more complicated systems including constraints one may wish to resort to a Hamilton--Dirac procedure in order to avoid any ambiguities, as we did for the scalar-tensor-spinor theory in the last subsection. It is actually instructive to compare the two approaches, so in the following we perform a canonical analysis of our toy model, which we hope will clarify the argument by removing the added complications related to general covariance and local Lorentz invariance that we had to deal with before. To this end we first write the action \eqref{eq:toy model action} in first order form with the help of a new variable $\psi$ and a Lagrange multiplier $\lambda$, both being bosonic ({\it i.e.} Grassmann-even)
\beq
S=\int dt\left[\phi\psi+\frac{1}{2}\,\chi\dot{\chi}+\frac{1}{2}\,\xi\dot{\xi}+g\chi\xi\phi\dot{\psi}+\lambda(\psi-\dot{\pi})\right]\,.
\eeq
Computing the canonical momenta (denoted as $p_\alpha$ for the variable $\alpha$) we find that these verify ${\cal C}_i = 0$, with ${\cal C}_i,\, i =1,\ldots,6$, defined by 
\beq\begin{gathered}
{\cal C}_1 = p_{\phi}\,,\qquad 
{\cal C}_2 = p_{\chi}-\frac{1}{2}\chi \,,\qquad 
{\cal C}_3 =  p_{\xi}-\frac{1}{2}\xi  \,,\qquad \\
{\cal C}_4 = p_{\psi}- g\chi\xi\phi\  \,,\qquad 
{\cal C}_5 = p_\pi + \lambda \,,\qquad 
{\cal C}_6 =  p_{\lambda}\,
\end{gathered}\eeq
which yields 6 primary constraints ${\cal C}_i \approx 0$, among which ${\cal C}_2$ and ${\cal C}_3$ are Grassmann-odd while the other are Grassmann-even. Note that we have kept here the Lagrange multiplier $\lambda$ dynamical given the dynamical analogy between $\lambda$ and $\phi$, both of which have vanishing momenta. 
We get hence the total Hamiltonian (see e.g.\ \cite{Henneaux:1992ig})
\beq
H_{\rm T}=-\phi\psi - \lambda \psi + \sum_{i=1}^{i=6} \gamma_i {\cal C}_i 
\eeq
with the $\gamma_i$ enforcing the primary constraints, two of them ($\gamma_2$ and $\gamma_3$) being Grassmann-odd while the other are Grassmann-even. 
Note that the Lagrangian being linear into the first derivatives of the fields, the canonical Hamiltonian is the pure potential $-(\phi+\lambda) \psi$. 

The next step is to check the time preservation of the primary contraints by computing $\{{\cal C}_i,H_{\rm T}\}$:
\ba
\{{\cal C}_1,H_{\rm T}\} = 0 &\Rightarrow& {\cal D}_1 \equiv \psi+ \gamma_4 g \chi\xi=0\,, \label{D10}\\
\{{\cal C}_2,H_{\rm T}\} = 0 &\Rightarrow& {\cal D}_2 \equiv \gamma_2 - \gamma_4 g \phi \xi =0 \,,\label{D20}\\
\{{\cal C}_3,H_{\rm T}\} = 0 &\Rightarrow& {\cal D}_3 \equiv \gamma_3 + \gamma_4 g \phi \chi =0 \,,\label{D30}\\
\{{\cal C}_4,H_{\rm T}\} = 0 &\Rightarrow& \phi+\lambda + g \chi \phi \gamma_3 - g \xi \phi \gamma_2 - g \chi \xi \gamma_1  =0 \,,\label{D40}\\
\{{\cal C}_5,H_{\rm T}\} = 0 &\Rightarrow& {\cal D}_5 \equiv \gamma_6 =0 \,,\label{D50}\\
\{{\cal C}_6,H_{\rm T}\} = 0 &\Rightarrow& {\cal D}_6 \equiv\psi-\gamma_5 = 0 \label{D60}
\ea
where we note in particular that the second and third lines above imply a simplification of the expression in the fourth, indeed e.g.\ the preservation of ${\cal C}_2$ implies that $\gamma_2 = \gamma_4 g \phi \xi$ which when inserted into the equation obtained by writing the preservation of ${\cal C}_4$ make the term depending on $\gamma_2$ vanish because $\xi$ squares to zero (and a similar reasoning holds for the term depending on $\gamma_3$). Hence the time preservation of ${\cal C}_4$ yields simply 
\ba
\{{\cal C}_4,H_{\rm T}\} = 0 &\Rightarrow& {\cal D}_4 \equiv \phi+\lambda  - g \chi \xi \gamma_1  =0 \,.
\ea
We thus have 6 equations ${\cal D}_i=0$  generated by the time preservation of the 6 primary constraints ${\cal C}_i=0$ and 6 Lagrange multipliers $\gamma_i$. If we would not work with supernumbers, a simple examination of these equations would lead to the conclusion that the process would stop here as the above equation would determine fully all the $\gamma_i$, and we would conclude that the number of propagating DoF would be 6 (in the Hamiltonian sense) in agreement with the expectation that higher order equations would result in an increase of the number of propagating DoF. However, we deal with supernumbers here and we should carefully level-decompose the above ${\cal D}_i$. While the Lagrange multipliers $\gamma_6$ and $\gamma_5$ are fully determined by equations (\ref{D50}) and (\ref{D60}) and that the (Grassmann-odd) Lagrange multipliers  $\gamma_2$ and $\gamma_3$ are fully determined once $\gamma_4$ is known using equations (\ref{D20}) and (\ref{D30}), we see that the vanishing of the bodies of ${\cal D}_1$ and ${\cal D}_4$ yield the additional constraints  
\ba 
\psi^{(0)} = 0 \label{psi0} \,,\\
\phi^{(0)} + \lambda^{(0)} =0 \label{phi0}
\ea
and the recursion relation for the Lagrange multipliers $\gamma_4$ and $\gamma_1$
\ba
\psi^{(2n)} + g \sum_{k=1}^{k=n}\gamma_4^{(2n-2k)}\left(\chi \xi\right)^{(2k)},  \label{recur4} \\
\phi^{(2n)} + \lambda^{(2n)} + g \sum_{k=1}^{k=n}\gamma_1^{(2n-2k)}\left(\chi \xi\right)^{(2k)}. \label{recur1}
\ea
Now if we compute the time evolution of $\psi$ and $\phi+\lambda$ we get 
\ba
\{\psi,H_{\rm T}\} = \gamma_4 ,\\
\{\phi + \lambda, H_{\rm T}\} =\gamma_1 + \gamma_6.
\ea
This together with (\ref{psi0}), (\ref{phi0}) and (\ref{D50}) yields in turn $\gamma_4^{(0)}=0$ and $\gamma_1^{(0)}=0$. Using then these equalities, the above two equations and the recursion relations (\ref{recur4}) and (\ref{recur1}) we conclude that $\gamma_4$ and $\gamma_1$ as well as $\psi$ and $\phi+ \lambda$ vanish at all levels. These latter two expressions being secondary constraints, which obviously do not generate any tertiary constraints. So we have a total of 8 (second class) constraints for 6 canonical pairs of variables, giving the expected result of $4$ mechanical DoF that we obtained previously.


\section{Discussion} \label{sec:discussion}

We have set out in this work to perform a first general analysis on the consistency of matter coupling in generalized scalar-tensor theories of the degenerate type. Our essential criterion that determined whether a given matter field can be described consistently within DHOST was that the interaction between the matter and scalar-tensor sectors that derives from the minimal coupling prescription should not introduce extra degrees of freedom. The possibility of having this issue is nicely illustrated by the simple example of veiled gravity minimally coupled to an ordinary scalar field. More generally, we have seen that, if such a pathology is present, it manifests itself in the Hamiltonian language through a loss of constraints, and we have explained by means of two examples the precise ways in which this can happen.

The first case occurs when the minimal coupling results in a mixing between the time derivatives of the metric and matter fields so that the full kinetic matrix does not have a block-diagonal form, implying that its rank may differ from the number of primary constraints one had before introducing the coupling. Although our example model of a non-canonical vector matter field was rather artificial, theories that involve a kinetic mixing with gravity are not hard to find. For instance any bosonic higher-spin theory as well as non-minimally coupled lower spin matter are dangerous in this sense. It should be remembered, however, that such an issue is not specific to DHOST theories and is already problematic in pure GR; this was our motivation to consider the artificial vector model, which is consistent in GR according to the aforementioned criterion.

The second possibility is that the gravitational coupling preserves the primary constraints of the scalar-tensor and matter sectors, but that it implies the failure to generate the necessary secondary constraints for the correct counting of DoF. This situation is more interesting as it does not occur in pure GR (which has only first class constraints), but is instead generic of degenerate modified gravity theories. Indeed, whenever a degeneracy is present, if the associated primary constraint fails to commute with any of the constraints in the matter sector then its minimal coupling to gravity must be deemed inconsistent.

This last remark motivates a careful study of spinor fields in the context of DHOST, which we have undertaken here by performing a full Hamilton--Dirac analysis of a minimally coupled Majorana spinor. We have shown that the required secondary constraints are actually present so that the consistency criterion is satisfied, although we have also pointed out that this property was not specific to the structure of the DHOST action. Rather, the commutation of the constraints is almost immediate once the level decomposition of the Grassmann algebra is taken into account. Even though this observation is somewhat trivial, it does lead to the perhaps unappreciated possibility of having a very large class of higher-derivative operators within modified gravity without introducing extra pathological DoF classically. We already mentioned, as an example, that the curvature-dependent spinor ``mass'' terms of the form $R^n\bar{\lambda}\lambda$ (some of which have been considered recently in \cite{Struckmeier:2018psp,Benisty:2019jqz}) can be seen to be harmless from the point of view of the DoF counting. There is however a non-trivial aspect in our result, namely that the coupling to the spinor matter field does not modify the form of the DHOST constraint. We have moreover shown, through the results of the appendix, that this crucial property is not an accident of the simple quadratic spinor model that we focused on in the main text, but that it actually holds for a very general set of self-interacting spin-1/2 theories. That being said, there are several interesting fermionic models that are not covered by our results, so we hope to revisit this question in future work. These include the cases of a spin-3/2 fermion, of multiple spinors involving mutual interactions, and of generalized spinor theories with actions that are not simply linear in the derivative of the field.

It is worth emphasizing again that our criterion on the coupling to matter is only a first requirement for the overall consistency of the DHOST framework and that phenomenological considerations may certainly impose additional constraints on the space of viable matter-coupled theories. However, these considerations depend on the specific applications of the models considered, in contrast to our analysis which is based on formal criteria, and an analysis of phenomenological factors lies outside the scope of this work. The current and main application of DHOST theories is to address cosmic acceleration, however, similar to vanilla scalar-tensor theories, one can consider many non-cosmological applications and our analysis will apply there as well. Nonetheless, it is an intriguing question whether our results can potentially complement the existing constraints on such cosmological applications, in particular the recent results on the decay of gravitational waves into dark energy \cite{Creminelli:2018xsv}, on the destabilization of dark energy inhomogeneities by gravitational waves \cite{Creminelli:2019kjy}, and on the Vainshtein screening mechanism \cite{Crisostomi:2019yfo} (see also \cite{Hirano:2019scf,Creminelli:2019nok,Noller:2020afd,Anson:2020fum} and below for other related works). At present these results are quite orthogonal to our work given that the exotic types of matter that we have so far identified as problematic do not play any role in these analyses. Nevertheless, we can foresee potentially useful applications in the context of violent astrophysical events, such as the merger of two neutron stars, in which higher-derivative corrections to the matter sector could in principle become important, while the Planck-suppressed corrections to the DHOST sector still be small. This theory would thus effectively be described by DHOST gravity coupled to matter fields that exhibit higher-derivative interactions, which may therefore spoil the degeneracy condition according to our results, under the assumption that the gravity sector can be treated classically in this regime.

Another interesting prospect would be to generalize our analysis by allowing for non-minimal coupling, although this is likely to require a case by case study. In fact, since there is a subset DHOST models that are related to non-degenerate theories via field redefinitions, it is clear that any matter field can be accommodated within this class through a non-minimal coupling if it can be covariantized consistently in the non-degenerate case. Another intriguing avenue to pursue is the study of matter coupling within the Palatini formalism recently developed for Horndeski and DHOST theories in \cite{Helpin:2019kcq}, as this is a very natural setting to consider the gravitational interaction with fermions. Finally, perhaps the most important restriction in our analysis was the assumption that matter does not couple directly to the DHOST scalar field, while for physical applications such a coupling might in fact be necessary in order to achieve non-trivial fifth force effects possibly screened {\it \`a la} Vainshtein \cite{Vainshtein:1972sx,Deffayet:2001uk,Babichev:2009jt,Babichev:2010jd} (see also e.g.\ \cite{Babichev:2009ee,Babichev:2013usa,Berezhiani:2013dw}). It is therefore an interesting problem to understand how a coupling of the form $\phi T^{\mu}_{\phantom{\mu}\mu}$, but also more general ones, can affect our conclusions.

\begin{acknowledgments}

We are grateful to T.\ Damour, C.\ de Rham, A.\ Vainshtein and P.\ Vanhove for some helpful conversations and comments. The authors acknowledge support by the European Research Council under the European Community's Seventh Framework Programme (FP7/2007-2013 Grant Agreement no.\ 307934, NIRG project) and by the European Union's Horizon 2020 Research Council grant 724659 MassiveCosmo ERC-2016-COG.

\end{acknowledgments}


\appendix

\section{General spinor action} \label{sec:appendix}

In this appendix we prove that the most general action for a Majorana spinor $\lambda$ involving at most a single derivative of the field takes a very simple form, one which ensures the consistency of minimal gravitational coupling both in GR and in DHOST.\footnote{In order to lighten the notation in this appendix we will omit spinor indices, so that a spinor bilinear is written for instance as $\bar{\lambda}\Gamma\lambda$, where $\bar{\lambda}$ is the Majorana conjugate of $\lambda$ and $\Gamma$ is any product of gamma matrices.}

To this end we begin by classifying all the independent products of spinor bilinears under the above assumption on the total number of derivatives. We recall that in four dimensions the set $\{\mathbf{1},\gamma^a,\gamma^{ab},\gamma^a\gamma_5,\gamma_5\}$ provides a complete basis of the complex $4\times4$ matrices. Note however that due to the Majorana condition only the bilinears $\bar{\lambda}\lambda$, $\bar{\lambda}\gamma_5\lambda$ and $\bar{\lambda}\gamma^a\gamma_5\lambda$ are non-zero among the candidates without derivatives \cite{Freedman:2012zz}. We therefore have four possible Lorentz invariant products of bilinears that have no derivatives:
\beq\begin{gathered}
A_1=(\bar{\lambda}\lambda)^2\,,\qquad A_2=(\bar{\lambda}\gamma_5\lambda)^2\,,\qquad A_3=\bar{\lambda}\gamma^a\gamma_5\lambda\cdot \bar{\lambda}\gamma_a\gamma_5\lambda\,,\\
B_1=\bar{\lambda}\lambda\cdot \bar{\lambda}\gamma_5\lambda\,.
\end{gathered}\eeq
These terms are not all independent due to Fierz identities. The product $B_1$ transforms into itself under a Fierz rearrangement, which can be used to show that it vanishes. The $A_n$ terms are reshuffled among themselves upon Fierzing and from the resulting relations it is straightforward to show that
\beq
A_2=A_1\,,\qquad A_3=-4A_1\,,\qquad B_1=0\,,
\eeq
so that there is a single independent non-trivial product, that we take to be $A_1$. Products of bilinears containing one derivative of the spinor field can be studied in the same way. We now count eight Lorentz invariant structures:\footnote{It may be thought that the term $C_4$ is redundant as it involves the matrix $\gamma^{ab}\gamma_5$, which is not part of the basis we have chosen. However, using a familiar identity, we can rewrite the product as $C_4=\frac{i}{2}\,\epsilon^{abcd}\bar{\lambda}\gamma_a\gamma_5\lambda\cdot \bar{\lambda}\gamma_{bc}\partial_d\lambda\,$.}
\beq\begin{gathered}
C_1=\bar{\lambda}\lambda\cdot \bar{\lambda}\gamma^a\partial_a\lambda\,,\qquad C_2=\bar{\lambda}\gamma_5\lambda\cdot \bar{\lambda}\gamma^a\gamma_5\partial_a\lambda\,,\\
C_3=\bar{\lambda}\gamma^a\gamma_5\lambda\cdot \bar{\lambda}\gamma_5\partial_a\lambda\,,\qquad C_4=\bar{\lambda}\gamma_a\gamma_5\lambda\cdot \bar{\lambda}\gamma^{ab}\gamma_5\partial_b\lambda\,,\\
D_1=\bar{\lambda}\gamma_5\lambda\cdot \bar{\lambda}\gamma^a\partial_a\lambda\,,\qquad D_2=\bar{\lambda}\lambda\cdot \bar{\lambda}\gamma^a\gamma_5\partial_a\lambda\,,\\
D_3=\bar{\lambda}\gamma^a\gamma_5\lambda\cdot \bar{\lambda}\partial_a\lambda\,,\qquad D_4=\bar{\lambda}\gamma_a\gamma_5\lambda\cdot \bar{\lambda}\gamma^{ab}\partial_b\lambda\,.\\
\end{gathered}\eeq
It is easy to see that the $C_n$ and $D_n$ groups transform independently of each other under Fierzing. From the resulting identities we find that only one product in each group is non-redundant, which we take to be $C_1$ and $D_1$,
\beq\begin{gathered}
C_2=-C_1\,,\qquad C_3=C_1\,,\qquad C_4=3C_1\,,\\
D_2=D_1\,,\qquad D_3=-D_1\,,\qquad D_4=-3D_1\,.
\end{gathered}\eeq
Lastly we also have to consider higher degree Lorentz invariant contractions of spinor bilinears. For purely potential terms without derivatives it is clear from the above result that all such terms are simply powers of the bilinear $\bar{\lambda}\lambda$. For the terms with one derivative we have only two new candidate structures given by
\beq
E_1=\bar{\lambda}\gamma^a\gamma_5\lambda\cdot \bar{\lambda}\gamma^b\gamma_5\lambda\cdot \bar{\lambda}\gamma_a\partial_b\lambda\,,\qquad E_2=\bar{\lambda}\gamma^a\gamma_5\lambda\cdot \bar{\lambda}\gamma^b\gamma_5\lambda\cdot \bar{\lambda}\gamma_a\gamma_5\partial_b\lambda\,.
\eeq
However the Fierz identity
\beq
\bar{\lambda}\gamma^a\gamma_5\lambda\cdot \bar{\lambda}\gamma^b\gamma_5\lambda=-\frac{2}{3}\,\eta^{ab}(\bar{\lambda}\lambda)^2\,,
\eeq
allows us to conclude that the $E_n$ products are not independent from the ones we have already classified.

From these results we infer that the most general action for a single Majorana spinor in flat space and including at most one derivative is given by
\beq \label{eq:general spinor action}
S_m=\int d^4x\Big\{\left[P_1(\bar{\lambda}\lambda)+c_1\bar{\lambda}\gamma_5\lambda\right]\bar{\lambda}\gamma^a\partial_a\lambda+P_2(\bar{\lambda}\lambda)\Big\}\,,
\eeq
where $P_1$ and $P_2$ are any (real) entire functions and $c_1$ is an arbitrary coupling constant.

Next we wish to show that upon covariantization the action \eqref{eq:general spinor action} does not thwart the constraints that ensure the consistency of the coupling to DHOST gravity. The existence of a spinor constraint is manifest from the fact that the action is linear in the time derivative of the field. Explicitly, the constraint \eqref{eq:spinor constraint} is modified in the presence of self-interactions as
\beq
\Lambda^{\alpha}:=\varpi^\alpha-\sqrt{\gamma}\left[P_1(\bar{\lambda}\lambda)+c_1\bar{\lambda}\gamma_5\lambda\right]\lambda^{\beta}(E^0)_{\beta}^{\phantom{\beta}\alpha}\approx0\,.
\eeq
On the other hand, the contribution of the matter action to the canonical momentum conjugate to the tetrad is also modified,
\beq
\pi_{(m)a}^{\phantom{(m)a}i}=-\frac{1}{4}\,\sqrt{\gamma}\left[P_1(\bar{\lambda}\lambda)+c_1\bar{\lambda}\gamma_5\lambda\right]\lambda^{\alpha}(\gamma_aE^iE^0)_{\alpha}^{\phantom{\alpha}\beta}\lambda_{\beta}\,,
\eeq
but it remains true that
\beq
\pi_{(m)}^{ij}:=\frac{1}{2}\,e^{a(i}\pi_{(m)a}^{\phantom{(m)a}j)}=0\,.
\eeq
The last relation guarantees that the DHOST constraint is unmodified by the presence of the spinor field. This completes the proof of the consistency of matter coupling for spinorial matter with general self-interactions.


\bibliographystyle{apsrev4-1}
\bibliography{DHOSTmatterBiblio}

\end{document}